\documentclass[reprint,amsmath,amssymb,aps,prc]{revtex4-2}

\usepackage{graphicx}
\usepackage{dcolumn}
\usepackage{bm}
\usepackage{xcolor}

\begin{document} 

\title {The Effects of Multi-$\Lambda$ Hyperons on Collective Modes in Nuclei}

\author{Bahruz Suleymanli}
\affiliation{Physics Department, Yildiz Technical University, 34220 Esenler, Istanbul, Türkiye}

\author{Kutsal Bozkurt}
\affiliation{Physics Department, Yildiz Technical University, 34220 Esenler, Istanbul, Türkiye}
\affiliation{Universit\'e Paris-Saclay, CNRS/IN2P3, IJCLab, 91405 Orsay, France}

\author{Elias Khan}
\affiliation{Universit\'e Paris-Saclay, CNRS/IN2P3, IJCLab, 91405 Orsay, France}
\affiliation{Institut Universitaire de France (IUF)}

\author{Ha\c{s}im G\"{u}ven}
\affiliation{Faculty of Engineering and Architecture, Istanbul Esenyurt University, 34510 Esenyurt, Istanbul, Türkiye}
\affiliation{Universit\'e Paris-Saclay, CNRS/IN2P3, IJCLab, 91405 Orsay, France}

\author{J\'er\^ome Margueron}
\affiliation{International Research Laboratory on Nuclear Physics and Astrophysics, Michigan State University and CNRS, East Lansing, MI 48824, USA}

\date{\today}

\begin{abstract}

The dynamical influence of $\Lambda$ hyperons on the excited-state properties of closed-shell 
multi-$\Lambda$ Ca, Ni, Sn and Pb hypernuclei is investigated using the self-consistent Hartree-Fock + Random 
Phase Approximation in coordinate space. The strength distributions for the isoscalar monopole, isovector 
dipole, and isoscalar quadrupole modes are calculated, revealing a systematic upward energy shift with 
increasing $\Lambda$ hyperon number $-S$. The scaling behavior of the computed centroid energies 
$\sqrt{m_1/m_{-1}}$ with respect to both the mass and hyperon number is determined. The nuclear 
incompressibility modulus $K_A$ is found to increase monotonically with $-S$. The largest value is found 
in the $^{258}_{50\Lambda}$Pb hypernucleus, reaching $K_A = 322 $ MeV. Calculations in uniform
hypernuclear matter confirm that this stiffening is a bulk effect driven by both the $N\Lambda$ and $\Lambda\Lambda$ interactions.
Analysis of the transition densities for states with maximal collective coherence 
indicates that the dynamical effect of $\Lambda$ hyperons is predominantly in phase with the protons, 
especially in the case of the isovector E1 modes. 
\end{abstract}

\maketitle

\section{\label{sec:intro}Introduction}

One of the most challenging goals in modern nuclear physics is to understand the equation of state 
(EoS) of nuclear matter under extreme conditions, such as those found in the core of neutron stars \cite{Burgio2021, 
Lattimer2021, Chatziioannou2025}. The presence of $\Lambda$ hyperons, which differ from nucleons by their strangeness 
quantum number, is crucial for understanding these extreme environments. By extending the nuclear chart into the strangeness 
dimension, they provide unique insights into baryon-baryon interactions that remain inaccessible through standard nuclear 
scattering experiments \cite{Gal2016, Miwa2025}. To link these microscopic interactions to the macroscopic EoS, collective 
excitations involving $\Lambda$ hyperons are the ideal tool to probe global properties such as incompressibility and 
symmetry energy, parameters that are critical for modeling the stability of neutron stars against gravitational collapse.

To constrain the effective interactions governing these collective phenomena, one must rely on the precise 
spectroscopic data of hypernuclear excited states, which serve as the experimental benchmark for theoretical models.
While direct scattering is limited by the short lifetime of $\Lambda$ hyperons, high-resolution $\gamma-$ray spectroscopy
has successfully identified excited states in a wide range of single$-\Lambda$ hypernuclei ($A = 4\ \text{to}\ 19$)
\cite{HASHIMOTO2006, TAMURA2013, Sanchez2013, Feliciello2015, Tamura2020, Tamura2000, Ukai2004, Ukai2008, TAMURA2010,
Hosomi2015, J-PARC2015, Yang2018}. Notable examples include the resolution of fine-structure doublets and electromagnetic
transitions in $_\Lambda^7$Li \cite{SASAO2004, Tanida2001, TAMURA2000249, Ukai2006}, $_\Lambda^9$Be \cite{Akikawa2002}, and
$_\Lambda^{16}$O \cite{UKAI2005, ZHOU2005}. Furthermore, observations of transitions in $_\Lambda^{15}$N via proton 
emission demonstrate the capability to probe excited states in heavier daughter nuclei \cite{Ukai2008}. Beyond single$-
\Lambda$ systems, the experimental frontier is rapidly expanding toward multi-strange and heavier hypernuclei. Recent
advancements in machine-learning-based emulsion analysis are significantly increasing the detection efficiency of double$-
\Lambda$ hypernuclei \cite{Nakazawa2020, HE2025, Yoshida2021, Nakazawa2022, Nakazawa2024}, while collider experiments have
begun to report yields of (anti)hypernuclei like $_\Lambda^4$H and $_\Lambda^4$He that deviate from ground-state-only 
expectations, pointing to the significant role of excited states even in extreme environments \cite{ALICE2025}.

To date, the excited states of single- and double-$\Lambda$ hypernuclei have primarily been investigated 
through theoretical calculations based on mean-field approaches, see for instance Refs.~\cite{Cugnon2000,Vidana2001,
Margueron2017,Suleymanli2024} and references therein. Investigations employing Skyrme–Hartree–Fock 
(SHF) combined with random phase approximation (RPA) revealed that the introduction of two $\Lambda$ hyperons 
into nuclei such as $^{42}_{\Lambda\Lambda}$Ca, $^{122}_{\Lambda\Lambda}$Sn, and $^{210}_{\Lambda\Lambda}$Pb 
shifts isoscalar giant monopole resonance (ISGMR) strengths toward higher energies \cite{Lv2018}. 

Furthermore, dipole excitations in hypernuclei, specifically the soft dipole mode, exhibit unique characteristics 
due to the oscillation of a $\Lambda$ hyperon against the nuclear core. Using the sum-rule approach with Skyrme-type 
$\Lambda N$ interactions, a systematic study of hypernuclei ranging from $^{16}_{\Lambda}$O to $^{208}_{\Lambda}$Pb 
demonstrated that the excitation energies of the soft dipole $\Lambda$ mode decrease with increasing mass number 
\cite{Minato2013}. This soft dipole mode can be parametrized effectively, showing significant mixing of configurations 
beyond simple 1p$_\Lambda$-1s$_\Lambda$ transitions.

Deformation and rotational characteristics are also impacted by the addition of $\Lambda$ particles. Relativistic 
mean-field (RMF) calculations indicated that the introduction of a $\Lambda$ hyperon into deformed nuclei could 
substantially alter nuclear shapes, sometimes reducing or even eliminating deformation, as observed in certain 
isotopes such as Si and Mg \cite{Minato2013_2}. Additionally, low-lying collective modes like dipole, quadrupole, 
and octupole excitations were studied via SHF+RPA methods, indicating that adding $\Lambda$ hyperons shifts the 
excitation energies upward, suggesting that $\Lambda$ particles significantly influence the single-particle and 
collective excitations in hypernuclei \cite{Minato2012}. 

Recent beyond-mean-field studies have been extended to specific hypernuclei to examine the interplay between
single-particle and collective degrees of freedom. For $^{21}_{\Lambda}$Ne, calculations using the generator coordinate 
method based on covariant density functional theory revealed that the low-lying negative-parity states exhibit strong 
mixing between different $\alpha$-cluster configurations and $\Lambda$ single-particle states ($\Lambda_s$ and $\Lambda_p$),
where the inclusion of octupole correlations significantly lowers the energy levels and changes electric multipole 
transition strengths \cite{Xia2023}. Similarly, beyond-mean-field analyses of $^{9}_{\Lambda}$Be and Ne isotopes
demonstrated that while magnetic moments remain relatively insensitive to the $\Lambda N$ interaction, the $E2$ transition
strengths are strongly affected \cite{Yao2024, Xue2024}. In the medium-mass nucleus $_\Lambda^{37}$Ar, beyond-mean-field
calculations demonstrated that the $\Lambda$ hyperon stabilizes superdeformed bands built on distinct configurations (s, p,
sd), thereby modifying rotational spectra and transition probabilities through configuration-dependent impurity effects
\cite{Cui2022}.

While several studies have explored how $\Lambda$ hyperons affect the structure and dynamics of finite nuclei,
it is important to understand their impact in baryonic matter at high density since their presence gives rise to the hyperon puzzle: hyperons soften the EoS enough to limit the mass-radius relation below the observational constraint of 2$M_\odot$. Several theoretical predictions have, therefore, suggested a phenomenological correction to the interaction, which provide the repulsion necessary to stiffen the EoS and match with the observational constraints. These mechanisms include enhanced vector-meson repulsion via SU(3) flavor symmetry \cite{Weissenborn2012}, the inclusion of strongly repulsive hyperonic three-body forces ($\Lambda N N$) \cite{Lonardoni2015}, and the high-density phase transition via a hadron-quark crossover \cite{Masuda2016}.

Since previous research predominantly considered scenarios with only one or two $\Lambda$ hyperons, it is timely
to extend the study of excited states to multi-$\Lambda$ hypernuclei. 
The present study marks the first 
attempt to systematically investigate collective modes in multi-$\Lambda$ hypernuclei, offering deeper insights into their structure 
and dynamics. Unlike prior studies employing the Skyrme interaction in the $\Lambda N$ channel, the present work 
uses the NSC97f interaction derived from Brueckner–Hartree–Fock calculations, accurately reproducing experimental 
binding energies of all single-$\Lambda$ hypernuclei across a wide mass range \cite{Cugnon2000, Vidana2001}. It is 
worth noting that analyses of p-shell hypernuclei using the NSC97f force set for A $\geq$ 10 were able to reproduce 
nearly all doublet spacing energies with an accuracy of 30~keV \cite{Millener2011}. Furthermore, the empirical force set 
EmpC is employed for the $\Lambda\Lambda$ channel, precisely calculating the experimental binding energy (approximately 
1 MeV) of the $^6_{\Lambda\Lambda}$He hypernucleus \cite{Margueron2017}. This comprehensive study provides the first 
systematic and mode-resolved analysis of how multi-$\Lambda$ hyperon admixture modifies the collective excitation 
spectra of finite nuclei, thereby advancing our understanding of strangeness effects on nuclear structure and the 
equation of state under extreme conditions.

The rest of this paper is organized as follows. In Sec.~\ref{sec:2}, we provide a brief overview of the HF-RPA approach 
applied to multi-$\Lambda$ hypernuclei. Sec.~\ref{sec:3} presents the isoscalar and isovector response results for the 
multi-$\Lambda$ hyperisotopic chains. Conclusions are given in Sec.~\ref{sec:concs}. 


\section{\label{sec:2} Hartree-Fock-Random Phase Approximation approach for multi-$\Lambda$ hypernuclei}

In this section, we address the extension of the Hartree-Fock Random Phase Approximation (HFRPA) to multi-$\Lambda$ 
hypernuclei assuming spherical symmetry. The calculations are restricted to systems with closed-shell nuclear cores 
and filled $\Lambda$ subshells, where pairing correlations and deformation effects can be neglected. 
For such systems, HFRPA enables the determination of both ground and excited-state properties. 
Since the focus of the present work is on the excited-state results, only several points 
concerning the ground-state calculations are highlighted. Further details on the ground-state treatment are provided 
in Ref.~\cite{Vidana2001,Margueron2017,Suleymanli2024}.

\subsection{Formalism}

\begin{figure}
\includegraphics[width=0.48\textwidth]{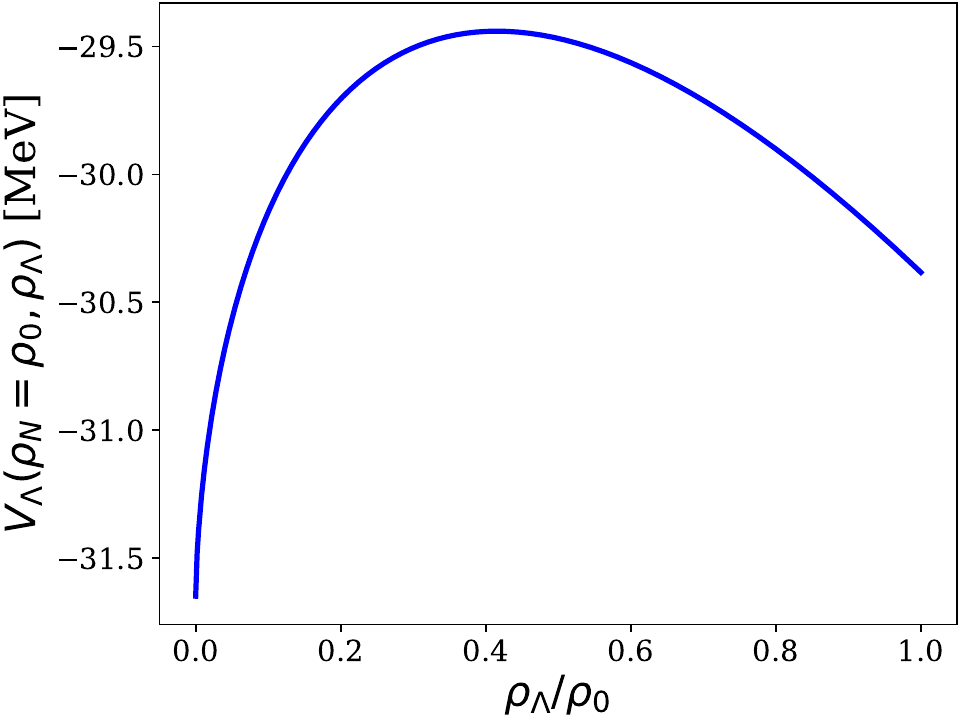}
\caption{$\Lambda$ mean-field potential $V_\Lambda$ as a function of
$\rho_\Lambda/\rho_0$ at fixed $\rho_N=\rho_0$ for the
SGII-NSC97f+EmpC functional.}
\label{fig:V_L}
\end{figure}

The total energy for hypernuclei within HF theory is computed using the energy-density functional (EDF) formalism as 
follows:
\begin{equation}
E = \int \!\!d^3 r \,\left(\sum_{i=\Lambda,N}\frac{\tau_i(r)}{2m_i^*\left(\rho_N(r)\right)} +\sum_{i,j=\Lambda,N}
s_{ij} \epsilon_{ij}(r)\right),
\label{eq:total_energy}
\end{equation}
where $\tau_i$ denote the kinetic energy densities associated with nucleons and $\Lambda$ hyperons, $m_i^*$ their 
corresponding effective masses, $s_{ij}$ is equal to $1$ when $i=j$ and $0.5$ when $i\neq j$, and the energy density 
terms $\epsilon_{ij}$ describe nucleon-nucleon ($NN$), hyperon-nucleon ($\Lambda N$), and 
hyperon-hyperon ($\Lambda \Lambda$) contributions.
Specifically, the $\Lambda N$ and $\Lambda \Lambda$ energy densities are defined as:
\begin{eqnarray}
\epsilon_{\Lambda\mathrm{N}} = \epsilon_{\mathrm{N}\Lambda} &=& -(\alpha_1-\alpha_2 \rho_\mathrm{N}+\alpha_3 
\rho_\mathrm{N}^2)\rho_\mathrm{N}  \rho_{\Lambda}\nonumber\\
&&+(\alpha_4-\alpha_5 \rho_\mathrm{N}+\alpha_6 \rho_\mathrm{N}^2) \rho_\mathrm{N} \rho_{\Lambda}^{5 / 3} \nonumber\\
&& - \left(\frac{m_{\Lambda}}{m_{\Lambda}^*\left(\rho_\mathrm{N}\right)}-1\right) \frac{3\left(3 \pi^2\right)
^{2 / 3}}{10 m_{\Lambda}} \rho_{\Lambda}^{5 / 3},
\label{eq:hyperon_nucleon}
\end{eqnarray}
\begin{equation}
\epsilon_{\Lambda \Lambda} = -(\alpha_7-\alpha_8 \rho_\Lambda+\alpha_9 \rho_\Lambda^2) 
\rho_{\Lambda}^2,
\label{eq:hyperon_hyperon}
\end{equation}
where 
\begin{equation}
\frac{m_{\Lambda}^*\left(\rho_\mathrm{N}\right)}{m_{\Lambda}}=\mu_1-\mu_2 \rho_\mathrm{N}+\mu_3 \rho_\mathrm{N}^2-\mu_4 
\rho_\mathrm{N}^3.
\label{eq:effective_mass}
\end{equation}
For the NSC97f+EmpC parametrization, the corresponding coefficients are
$\mu_1 = 0.93$, $\mu_2 = 2.19$ fm$^3$, $\mu_3 = 3.89$ fm$^6$, and $\mu_4 = 0$.
The nucleon-nucleon interaction component, $\epsilon_{NN}$, follows Skyrme energy-density form as in Refs.~\cite{Vidana2001,
Margueron2017,Suleymanli2024}. Minimizing the total energy functional with respect to single-particle wave functions 
yields the radial HF equation:
\begin{equation}\label{eq:HF_rad}
{\left[-\boldsymbol{\nabla} \cdot \frac{1}{2 m_q^*(r)} \boldsymbol{\nabla}+V_q(r)\right] \phi_q^i(r)} =-e_q^i \phi_q^i(r),
\end{equation}
where $q=n,p,\Lambda$ represents neutron, proton, and $\Lambda$ hyperon particles, respectively. The potentials $V_\Lambda$ 
and $V_N$ entering the HF equations are explicitly defined as:
\begin{eqnarray}
V_{\Lambda}&=&-(\alpha_1-\alpha_2 \rho_\mathrm{N}+\alpha_3 \rho_\mathrm{N}^2) \rho_
\mathrm{N} + \frac{5}{3}(\alpha_4-\alpha_5 \rho_\mathrm{N}+\alpha_6 \rho_\mathrm{N}^2) \nonumber\\ 
&&\hspace{-0.5cm} \times \rho_\mathrm{N} \rho_{\Lambda}^{2 / 3} -2\alpha_7\rho_{\Lambda}+3\alpha_8 \rho_\Lambda^2-4\alpha_9 
\rho_\Lambda^3 -\left(\frac{m_{\Lambda}}{m_{\Lambda}^*\left(\rho_\mathrm{N}\right)}-1\right)\nonumber\\
&&\hspace{-0.5cm} \times \frac{\left(3 \pi^2\right)
^{2 / 3}}{2 m_{\Lambda}} \rho_{\Lambda}^{2 / 3},
\label{eq:V_Lambda}
\end{eqnarray}
and
\begin{eqnarray}
V_N &=& -(\alpha_1-2\alpha_2 \rho_\mathrm{N}+3\alpha_3 \rho_\mathrm{N}^2) \rho_{\Lambda} +(\alpha_4-2\alpha_5 
\rho_\mathrm{N}+3\alpha_6 \rho_\mathrm{N}^2) \nonumber\\
&&\hspace{-0.5cm}\times \rho_{\Lambda}^{5 / 3} +\frac{\partial}{\partial \rho_\mathrm{N}}\left[\epsilon_\mathrm{NN} +
\left(\frac{m_{\Lambda}}{m_{\Lambda}^*\left(\rho_\mathrm{N}\right)}\right) \left(\frac{\tau_{\Lambda}}{2 m_{\Lambda}}
-\frac{3}{5} \frac{\left(3 \pi^2\right)^{2 / 3}}{2 m_{\Lambda}}\right.\right.\nonumber\\
&&\hspace{-0.5cm} \left.\left.
\times \rho_{\Lambda}^{5 / 3}\right)\right] -i W_N(r)(\boldsymbol{\nabla} \times \boldsymbol{\sigma})\,. 
\label{eq:V_N}
\end{eqnarray}
Fig.~\ref{fig:V_L} shows that $V_\Lambda$ remains negative at
$\rho_N=\rho_0$, where $\rho_0=0.16~{\rm fm}^{-3}$ is the nuclear saturation
density, confirming the attractive character of the $\Lambda$ mean field. The
dominant attractive contribution comes from the
$-(\alpha_1-\alpha_2\rho_N+\alpha_3\rho_N^2)\rho_N$ term, which is independent
of $\rho_\Lambda$ and equals $-31.65~{\rm MeV}$ at $\rho_N=\rho_0$ for the
SGII-NSC97f+EmpC calculation.

With the ground states established through these equations, the excited states of multi-$\Lambda$ hypernuclei can be 
systematically analyzed using the RPA. The standard RPA eigenvalue problem is formulated 
as:
\begin{equation}
\left(\begin{array}{cc}
A & B \\
-B & -A
\end{array}\right)\binom{X^{(\nu)}}{Y^{(\nu)}}=E_\nu\binom{X^{(\nu)}}{Y^{(\nu)}}
\label{eq:RPA}
\end{equation}
where $\nu$ indexes particle-hole configurations, $E_\nu$ is the excitation energy, and $X^{(\nu)}$ and $Y^{(\nu)}$ 
denote forward and backward amplitudes. The corresponding excitation operator for the RPA state $|\nu\rangle=O_\nu^{+}
|\tilde{0}\rangle$ is:
\begin{equation}
O_\nu^{+}=\sum_{m, i} X_{m i}^{(\nu)} c_m^+ c_i - Y_{m i}^{(\nu)} c_i^+ c_m,
\end{equation}
with $m,i$ denoting unoccupied and occupied states in the hypernuclear system, respectively.

The RPA matrices $A$ and $B$ are defined as:
\begin{equation}\label{eq:rpa_A}
A_{m i, n j}=\left(\epsilon_m-\epsilon_i\right) \delta_{m n} \delta_{i j}+\langle m j| V_\text{res}(N)+V_\text{res}
(\Lambda)|i n\rangle,
\end{equation}

\begin{equation}\label{eq:rpa_B}
B_{m i, n j}=\langle m n| V_\text{res}(N)+V_\text{res}(\Lambda)|i j\rangle.
\end{equation}
Here, $V_\text{res}(N)$ is associated with the residual interaction derived from $\epsilon_{NN}$ and includes the 
momentum-independent, momentum-dependent, spin–orbit, as well as Coulomb and Coulomb-exchange terms. The explicit  
forms of these terms can be found in Ref.~\cite{COLO2013}.   
The $\Lambda$-dependent residual interaction, $V_\text{res}(\Lambda)$, is defined as the second functional 
derivative of the total energy density with respect to the hyperon density:
\begin{equation}
V_\text{res}(\Lambda) = \frac{\delta^2 \left(\epsilon_{\Lambda \mathrm{N}} + \epsilon_{\Lambda \Lambda}\right)}{\delta \rho_{\Lambda}^2}.
\end{equation}
Upon performing the second derivative, terms linear in $\rho_\Lambda$ vanish, while the terms proportional to $\rho_\Lambda^{5/3}$ 
in Eq.~\eqref{eq:hyperon_nucleon} yield a dependence of $\rho_\Lambda^{-1/3}$. The resulting residual interaction is expressed as:
\begin{eqnarray}\label{eq:Vres_L}
V_\text{res}(\Lambda) &=&   \frac{10}{9} (\alpha_4 - \alpha_5 \rho_N + \alpha_6 \rho_N^2) \rho_N  \rho_\Lambda^{-1/3}\nonumber\\ 
&& -2\alpha_7 + 6\alpha_8 \rho_\Lambda -12\alpha_9 \rho_\Lambda^2\nonumber\\
&& -\left(\frac{m_{\Lambda}}{m_{\Lambda}^*\left(\rho_\mathrm{N}\right)}-1\right)\frac{\left(3 \pi^2\right)
^{2 / 3}}{3 m_{\Lambda}}\rho_\Lambda^{-1/3}.
\end{eqnarray}
 We note that, for all hypernuclei considered in the present work, the nucleon density $\rho_N$ decreases faster than $\rho_\Lambda^{1/3}$ in the
asymptotic region. Therefore, in the limit $\rho_\Lambda\to0$, Eq.~\eqref{eq:Vres_L} does not
diverge. We further note that, for a purely nucleonic system with $\rho_\Lambda\equiv0$, Eq.~\eqref{eq:Vres_L} does not enter the RPA calculation.

Finally, the isoscalar and isovector multipole response functions are
calculated from the corresponding transition operators. For a given magnetic
component $M$, the isoscalar operator is written as
\setcounter{equation}{13}
\begin{equation}
F_{JM}^{(IS)}
=
\sum_{i=1}^{A}
f_J(r_i)Y_{JM}(\mathbf{r}_i),
\label{eq:is}
\end{equation}
where $Y_{JM}(\mathbf{r}_i)$ denotes the spherical harmonic evaluated in the
angular direction of the coordinate $\mathbf{r}_i$. The isovector dipole operator is written with the hypernuclear center-of-mass
correction as
\begin{equation}
F_{1M}^{(IV)}
=
\sum_{i=1}^{Z}
\left[
r_{p,i}Y_{1M}(\mathbf{r}_{p,i})
-
R_{\rm cm}Y_{1M}(\mathbf{R}_{\rm cm})
\right].
\label{eq:iv}
\end{equation}
The center-of-mass coordinate entering Eq.~(\ref{eq:iv}) is given by
\begin{equation}
\mathbf{R}_{\rm cm}
=
\frac{
m_n\sum_{i=1}^{N}\mathbf{r}_{n,i}
+
m_p\sum_{i=1}^{Z}\mathbf{r}_{p,i}
+
m_\Lambda\sum_{i=1}^{-S}\mathbf{r}_{\Lambda,i}
}
{
m_nN+m_pZ-m_\Lambda S
}.
\label{eq:cm}
\end{equation}
In this notation, $N$, $Z$, and $-S$ denote the neutron, proton, and
$\Lambda$ hyperon numbers, respectively, while $A=N+Z-S$ is the total
baryon number. The total strength is obtained by summing over the magnetic
substates,
\begin{equation}
S_J(E)
=
\sum_{M=-J}^{J}
\sum_\nu
\left|
\left\langle \nu \left| F_{JM} \right| 0 \right\rangle
\right|^2
\delta(E-E_\nu).
\label{eq:strength}
\end{equation}

\subsection{Numerical details}

The ground state properties and excited state response functions were calculated using a fully self-consistent 
HF-RPA approach in coordinate space. The radial HF equations (\ref{eq:HF_rad}) were solved iteratively using the Numerov 
algorithm on a discretized radial mesh with a step size of 0.1 fm and a box radius of 20 fm, assuming spherical symmetry. 
Consequently, pairing correlations and deformation effects are neglected and calculations are restricted to systems with 
closed-shell nuclear cores and filled $\Lambda$ subshells. For the RPA calculations, the single-particle basis 
was constructed from the self-consistent HF mean field. The excited states were then determined by solving the RPA matrix 
equation (\ref{eq:RPA}). The configuration space included all occupied hole states and all unoccupied particle states up 
to an energy cutoff $E_c = 60$ MeV. This cutoff was chosen to ensure the convergence of the collective state properties 
and the exhaustion of the energy-weighted sum rules. The residual particle-hole interaction was derived consistently from 
the energy density functional, maintaining full self-consistency between the mean field ground state and the RPA excitations. In all 
cases, the RPA strength distributions were smeared using a Lorentzian function with a width parameter of 1~MeV.


\section{\label{sec:3} Results}

In this section, we present detailed results of our HFRPA calculations focusing on the multipole strength distributions 
for various multi-$\Lambda$ hypernuclei. The calculations are performed for Ca, Ni, Sn, and Pb hyper-isotopes, 
systematically varying the number of $\Lambda$ hyperons embedded within the nuclear medium. In both HF and RPA 
calculations, the Skyrme-type SGII interaction is employed in the $NN$ channel. It should be noted that 
SGII interaction is a well-calibrated Skyrme force that reproduces nuclear binding energies, radii, saturation properties, 
spin properties, and has been successfully applied to hypernuclear systems in which two $\Lambda$ hyperon states are 
considered in the GMR case, making it a reliable choice for the $NN$ interaction in $\Lambda$ hypernuclei studies \cite{Brenna2013, Lv2018}. In order to demonstrate that the obtained results do not qualitatively depend on the choice of the Skyrme interaction,
calculations of the multipole strength distributions for $^{208-S}_{-S\Lambda}$Pb hypernuclei are also performed with the SLy4 and SkM$^*$ interactions in the $NN$ channel.
For the hyperonic sector, the NSC97f interaction supplemented by the EmpC correction was adopted in the $\Lambda N$ and 
$\Lambda\Lambda$ channels, as this combination has been shown to provide a realistic description of hyper-nuclear binding 
and double-$\Lambda$ systems~\cite{Margueron2017}.

\subsection{The Isoscalar Monopole Response in Multi-$\Lambda$ Hypernuclei}

\begin{figure*}
\includegraphics[width=\textwidth]{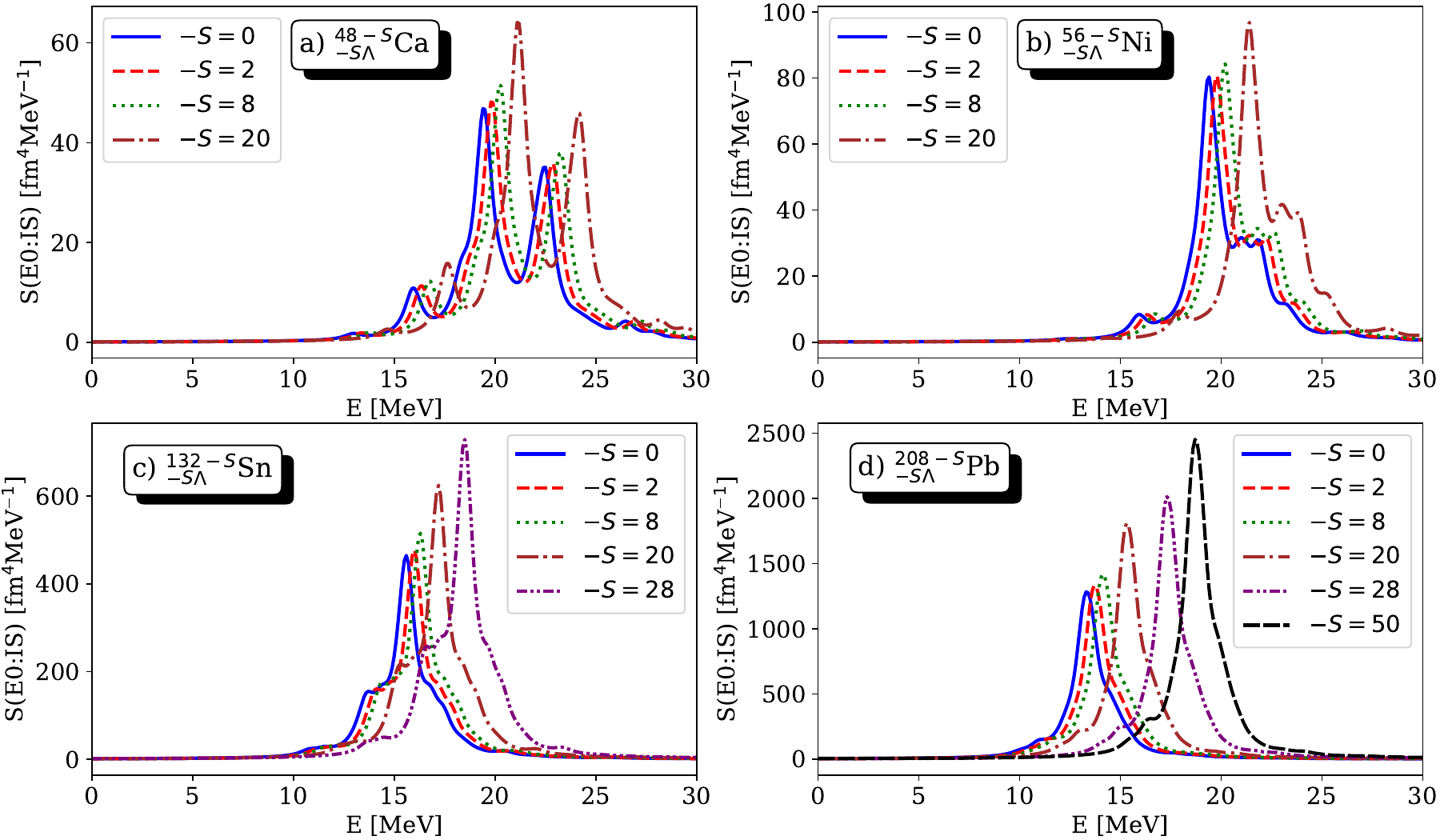}
\caption{Isoscalar monopole strength distributions for multi-$\Lambda$ hypernuclei $^{48-S}_{-S\Lambda}$Ca, $^{56-S}_ {-S\Lambda}$Ni, $^{132-S}_{-S\Lambda}$Sn, and $^{208-S}_{-S\Lambda}$Pb calculated within the HFRPA approach using the SGII-NSC97f+EmpC functional.}
\label{fig:monopole_full}
\end{figure*}

The isoscalar monopole 
strength
functions, $S(E0;IS)$, are depicted in the Fig.~\ref{fig:monopole_full} for excitation energies ranging from 0 to 30~MeV. 
In Fig.~\ref{fig:monopole_full}(a), cores made of $^{48}$Ca with different strangeness numbers ($-S$) are shown. The solid 
blue line corresponds to the nucleus without hyperons ($-S=0$). A pronounced peak near 20 MeV characterizes the monopole 
strength distribution, indicative of the primary monopole resonance. Introducing two $\Lambda$ particles 
($-S=2$, dashed red line) shifts this peak slightly toward higher energies, reflecting a shift up of the nuclear 
compressibility due to hyperon inclusion. Further increasing the hyperon number ($-S=8, -S=20$) progressively broadens 
the peak and systematically increases the excitation energies.

Similarly, Fig.~\ref{fig:monopole_full}(b) illustrates the results for $^{56}$Ni. The basic monopole resonance is 
positioned around 20 MeV for the non-strange nucleus. The introduction of hyperons, again shown for $-S=2, -S=8,$ 
and $-S=20$, exhibits clear shifts and enhancements in monopole strength, especially notable for the $-S=20$ 
configuration, demonstrating substantial hyperon-induced modifications in medium-mass nuclei.

For heavier nuclei shown in Fig.~\ref{fig:monopole_full}(c) and (d), the collective mode is less fragmented and mostly 
concentrated on a single peak. The height of the single peak is increased as a function of the strangeness number $-S$. 
For $^{132}$Sn shown in Fig.~\ref{fig:monopole_full}(c), the effects of hyperon inclusion are pronounced. With no hyperons, 
a substantial monopole peak appears near $16$ MeV. Increasing the 
hyperon count up to $-S=28$ prominently shifts and enhances the monopole distribution.
The largest effects appear in the heaviest nucleus studied, $^{208}$Pb (Fig.~\ref{fig:monopole_full}(d)). The 
non-strange nucleus displays a dominant monopole resonance around 14 MeV, reflecting its well-known compressional 
properties. Introducing hyperons progressively shifts the resonance significantly towards higher energies and notably 
amplifies the strength, especially evident at high strangeness numbers such as $-S=28$ and $-S=50$. These results 
vividly underline the strong influence of multiple $\Lambda$ hyperons on nuclear matter properties and their substantial 
modifications to nuclear excitations.

\subsection{The Isovector Dipole Response in Multi-$\Lambda$ Hypernuclei}

\begin{figure*}
\includegraphics[width=\textwidth]{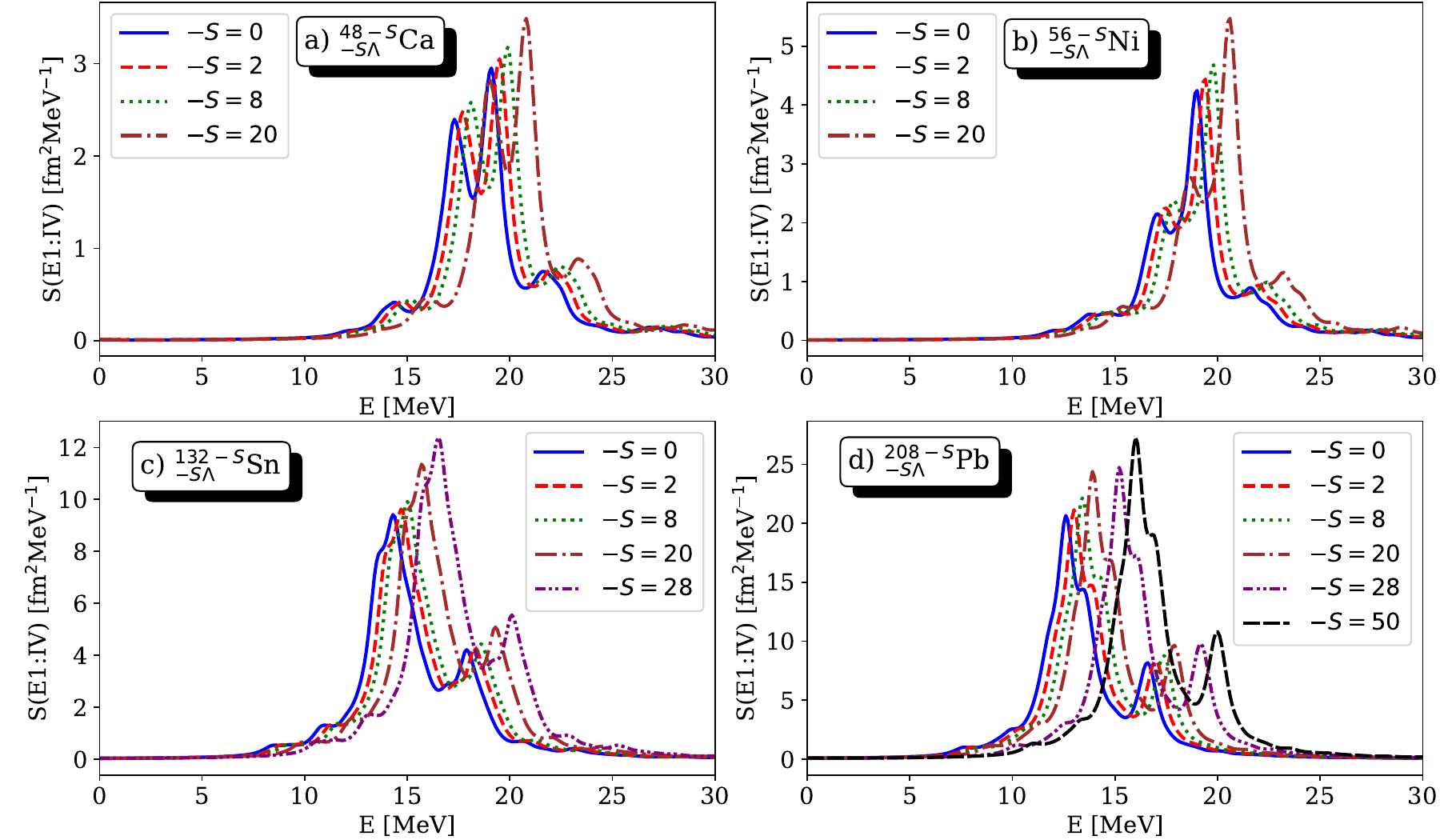}
\caption{Isovector dipole strength distributions for multi-$\Lambda$ hypernuclei $^{48-S}_{-S\Lambda}$Ca, 
$^{56-S}_ {-S\Lambda}$Ni, $^{132-S}_{-S\Lambda}$Sn, and $^{208-S}_{-S\Lambda}$Pb calculated within the HFRPA approach using the SGII-NSC97f+EmpC functional.}
\label{fig:dipole_full}
\end{figure*}

Fig.~\ref{fig:dipole_full} presents HFRPA predictions for the isovector dipole strength 
distributions $S(E1:IV)$ in multi-$\Lambda$ hypernuclei, specifically for the same set of nuclei ($^{48}$Ca, 
$^{56}$Ni, $^{132}$Sn, and $^{208}$Pb) and strangeness configurations analyzed in the monopole case 
(see Fig.~\ref{fig:monopole_full}).

For the core nucleus $^{48}$Ca ($-S=0$) shown in Fig.~\ref{fig:dipole_full}(a), the Giant Dipole Resonance (GDR) strength 
is concentrated in a broad peak centered around 18~MeV. This broad peak is mostly composed of two sub-peaks well visible 
in the figure. The addition of $\Lambda$ hyperons induces a systematic upward
shift of the entire resonance structure. As the strangeness content increases from $-S=2$ to $-S=20$, the peak
energy migrates towards higher energies, reaching approximately 20~MeV. A similar behavior is observed for $^{56}$Ni 
(Fig.~\ref{fig:dipole_full}b), where the main dipole peak at $\approx 17$~MeV shifts progressively to $\approx 19$~MeV 
at $-S=20$. In both light nuclei, the overall shape of the distribution is preserved, but the resonance is pushed to 
higher frequencies.

In the heavier systems, $^{132}$Sn and $^{208}$Pb (Figs.~\ref{fig:dipole_full}c and \ref{fig:dipole_full}d), the GDR 
exhibits a broad double-peak structure covering a wider energy window compared to lighter nuclei. The more pronounced 
peak is at lower energy than in lighter nuclei.
For $^{208}$Pb, the dominant peak is located at 14~MeV for $-S=0$. With the inclusion of hyperons, this peak shifts 
up significantly, reaching roughly 16~MeV for $-S=50$. 

Comparing the dipole results (Fig.~\ref{fig:dipole_full}) with the monopole response (Fig.~\ref{fig:monopole_full})
demonstrates that both collective modes exhibit a systematic shift upward with increasing strangeness. This indicates 
that the presence of $\Lambda$ hyperons impacts both the isoscalar and isovector sectors.
However, the mechanism differs. The monopole resonance is a breathing mode, and its shift reflects the
change in the curvature of the energy-density functional under compression. In
the present calculations, the density-dependent $N\Lambda$ and
$\Lambda\Lambda$ contributions increase the restoring force of the
isoscalar compression mode, leading to an upward shift of the monopole
resonance energy. The accompanying contraction of the nuclear density
distribution is a consequence of the additional attractive isoscalar mean field
generated by the $\Lambda$ hyperons (see Fig.~\ref{fig:V_L}).
In contrast, the dipole resonance is a translational oscillation of neutrons against protons. We observed and discussed previously that adding $\Lambda$ hyperons to the nuclear core modifies the equilibrium density distribution of the core, making it more compact. 
Thus, while the $\Lambda$ hyperons do not participate directly in the isovector oscillation, since they have isospin $T = 0$, they indirectly modify the restoring force and the spatial scale of the proton-neutron oscillation through the density rearrangement of the nuclear core, producing the upward shift of the GDR energy.

\subsection{The Isoscalar Quadrupole Response in Multi-$\Lambda$ Hypernuclei}

\begin{figure*}
\includegraphics[width=\textwidth]{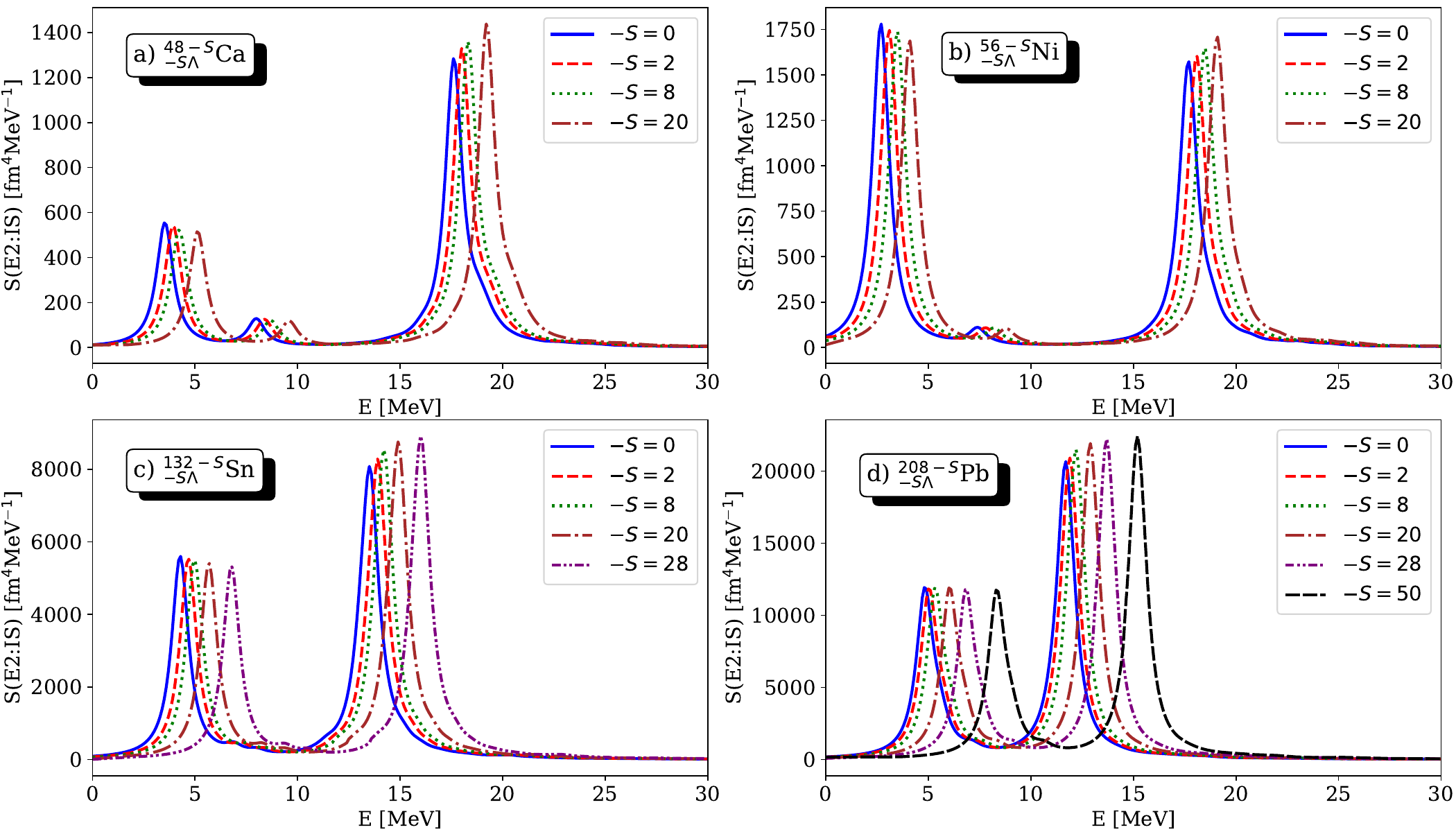}
\caption{Isoscalar quadrupole strength distributions for multi-$\Lambda$ hypernuclei $^{48-S}_{-S\Lambda}$Ca,
$^{56-S}_ {-S\Lambda}$Ni, $^{132-S}_{-S\Lambda}$Sn, and $^{208-S}_{-S\Lambda}$Pb calculated within the HFRPA approach using the SGII-NSC97f+EmpC functional.}
\label{fig:quadrupole_full}
\end{figure*}

Fig.~\ref{fig:quadrupole_full} displays the isoscalar quadrupole strength distributions (ISGQR) 
for multi-$\Lambda$ hypernuclei, again calculated within the HFRPA framework for the nuclei $^{48}$Ca, 
$^{56}$Ni, $^{132}$Sn, and $^{208}$Pb with varying strangeness numbers. The response is characterized by two distinct
regions: a low-energy state and a high-energy giant resonance peak.

For the light nuclei $^{48}$Ca and $^{56}$Ni (Figs.~\ref{fig:quadrupole_full}a and \ref{fig:quadrupole_full}b), the
strength distribution exhibits a prominent low-energy peak around $4-5$~MeV and a giant resonance structure near 
$17-18$~MeV. The addition of $\Lambda$ hyperons induces a clear and systematic upward shift in both regions. In $^{48}$Ca, the
high-energy peak shifts from approximately 17.5~MeV $(S=0)$ to nearly 19.5~MeV ($-S=20$). Similarly, in $^{56}$Ni, the
giant resonance moves from $\approx 17.5$~MeV to $\approx 19$~MeV. The low-energy states also exhibit a consistent,
though smaller, migration to higher energies.

In the heavier systems $^{132}$Sn and $^{208}$Pb (Figs.~\ref{fig:quadrupole_full}c and \ref{fig:quadrupole_full}d), the
GQR is dominated by a massive peak in the 10-15~MeV region. This collective mode is highly sensitive to the strangeness
content. In $^{208}$Pb, the peak energy increases from $\approx 11$~MeV for the core nucleus to over 13~MeV for the
hypernucleus with $-S=50$. This shift is accompanied by a preservation of the resonance width, indicating that the
$\Lambda$ hyperons stiffen the potential without significantly damping the collective motion.

A comparative analysis of the three collective modes reveals a unified physical picture. The ISGQR behaves qualitatively
similarly to the monopole and dipole resonances. All three modes exhibit a systematic hardening (increase in energy) with
increasing strangeness.
However, the magnitude of the shift in the quadrupole sector provides unique insights. The ISGQR is a surface mode,
involving the oscillation of the nuclear shape (from spherical to ellipsoidal). The observed energy increase indicates that the $\Lambda$ hyperons
generate an additional attractive isoscalar mean field, deepen the central
potential, and contract the nuclear density distribution. This density
rearrangement modifies the surface properties of the system and increases the
restoring force against quadrupole shape deformations. Unlike the dipole mode, which is purely isovector, the
isoscalar character of the quadrupole resonance allows for a more direct coupling to the $\Lambda$ hyperons through the
mean field, resulting in energy shifts that are comparable in magnitude to those observed in the monopole breathing
mode. Thus, the stiffening effect of strangeness is a global phenomenon, affecting volume (monopole), surface
(quadrupole), and translational (dipole) degrees of freedom.

\subsection{Sensitivity to the Skyrme interaction}\label{sec:A}

\begin{figure*}
\includegraphics[width=0.32\textwidth]{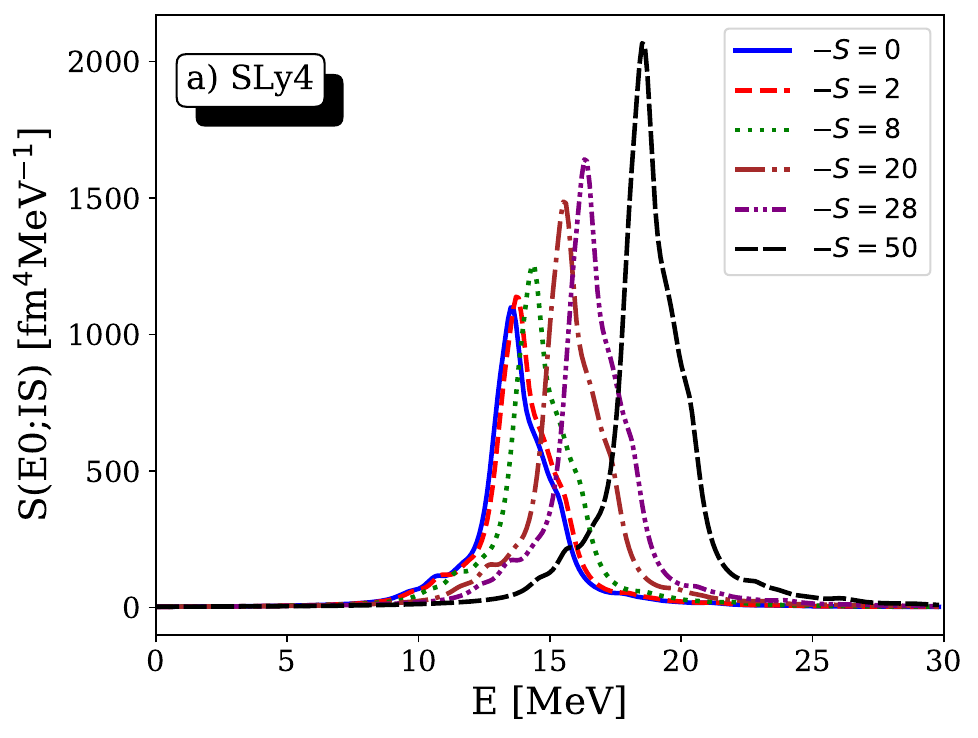}
\hfill
\includegraphics[width=0.32\textwidth]{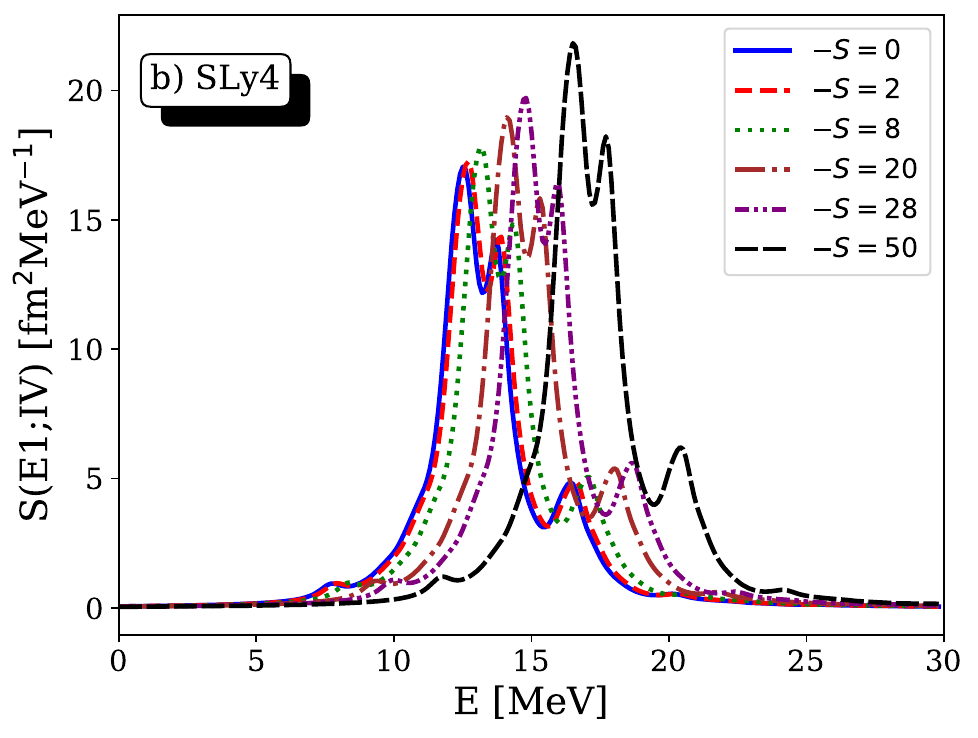}
\hfill
\includegraphics[width=0.32\textwidth]{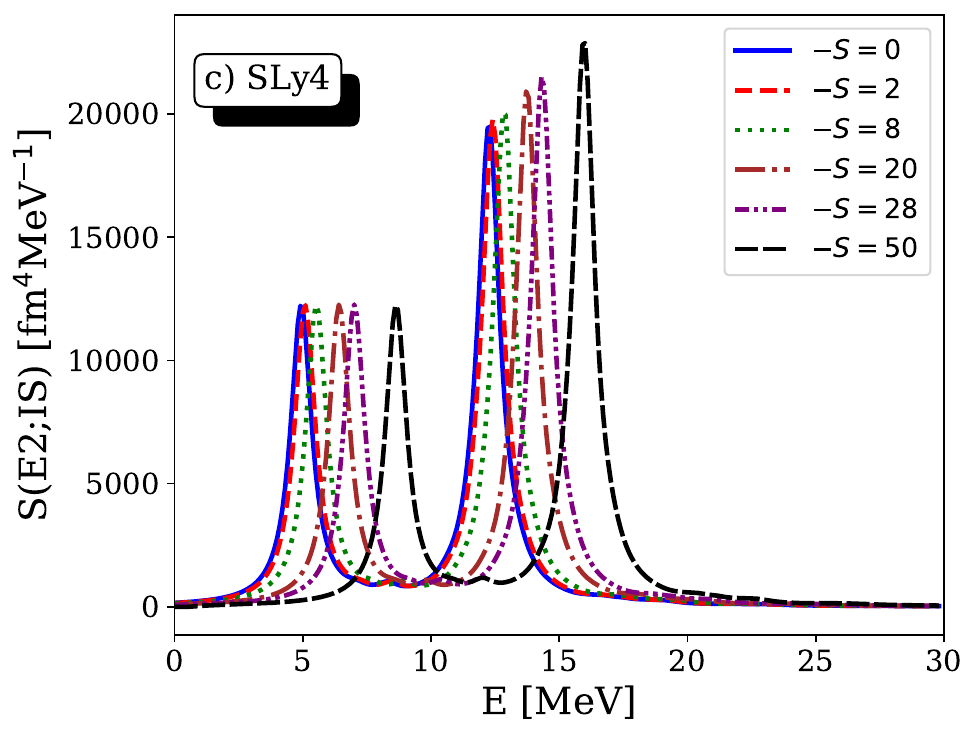}
\hfill
\includegraphics[width=0.32\textwidth]{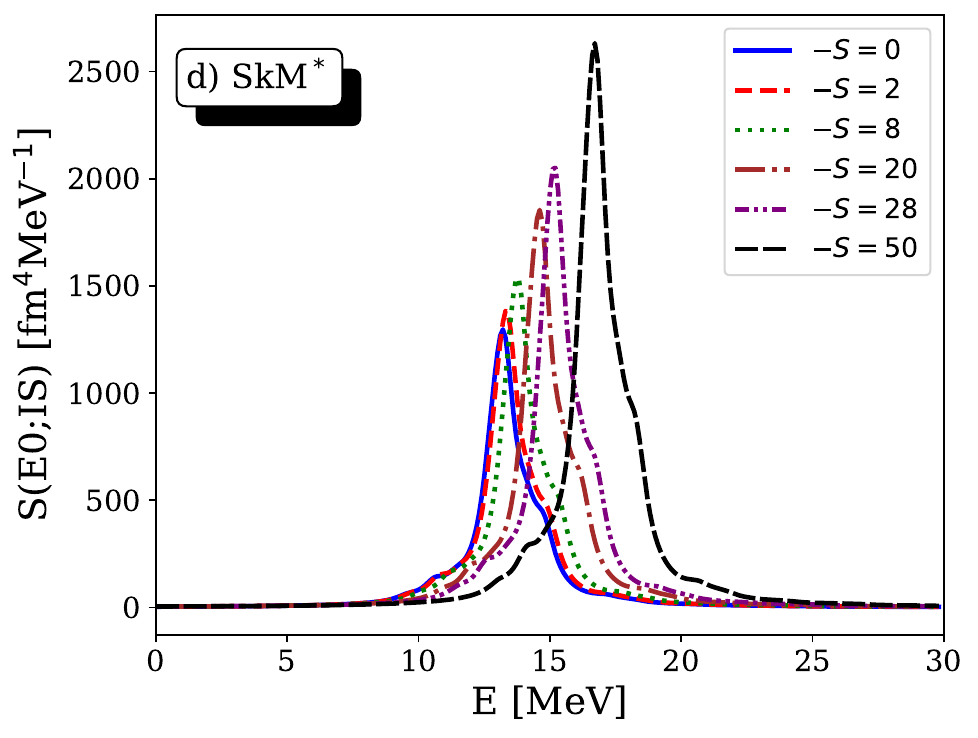}
\hfill
\includegraphics[width=0.32\textwidth]{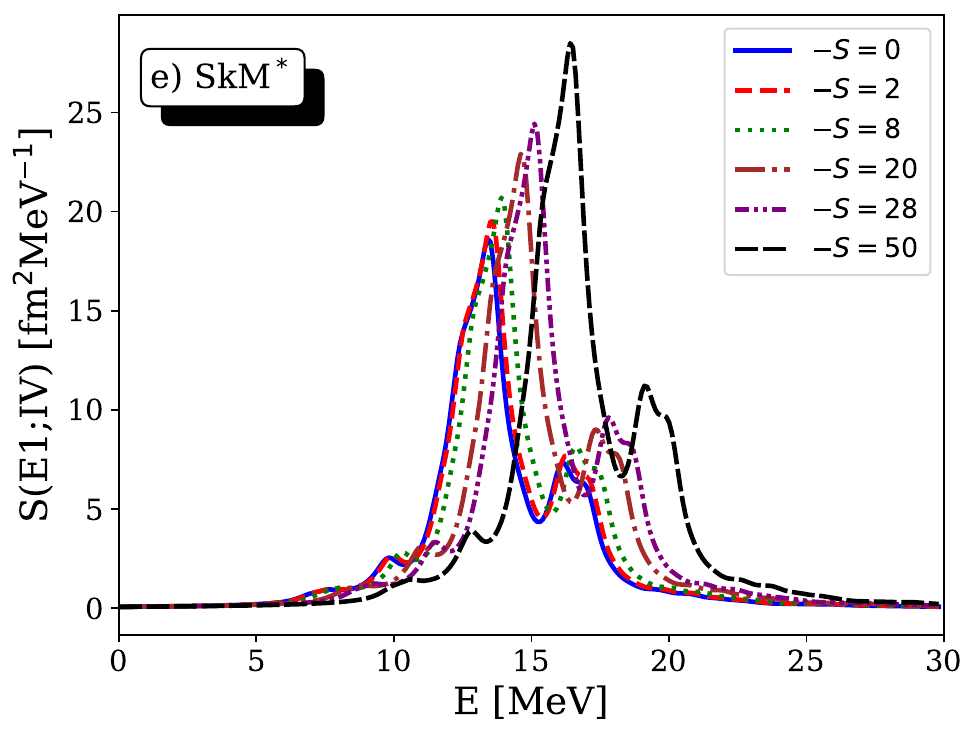}
\hfill
\includegraphics[width=0.32\textwidth]{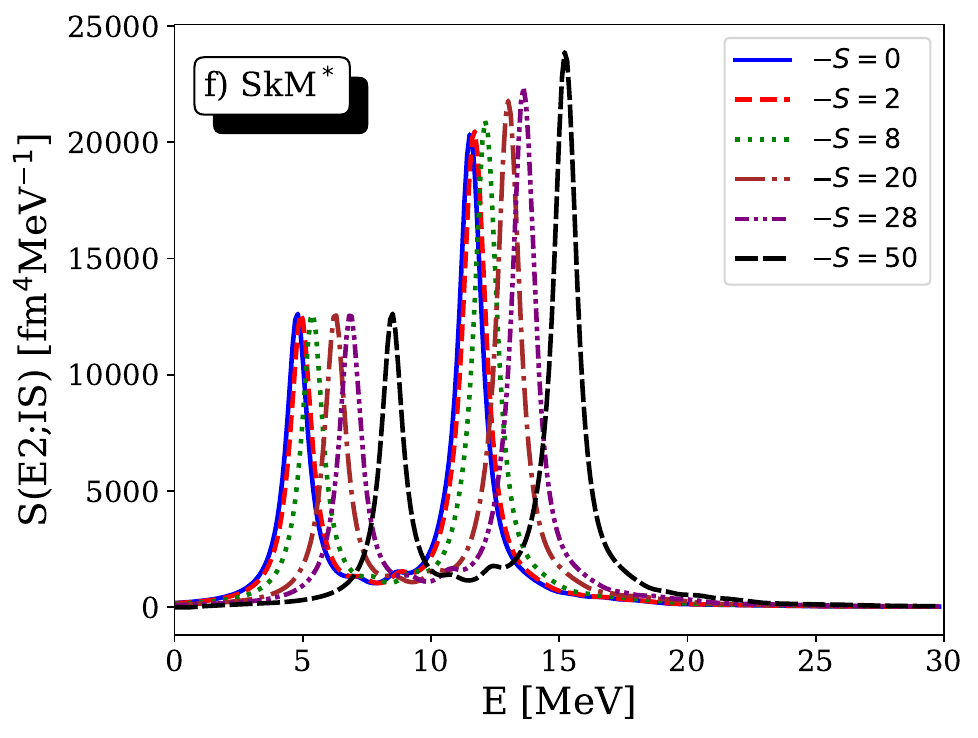}
\caption{Multipole strength distributions for $^{208-S}_{-S\Lambda}$Pb hypernuclei, obtained with the SLy4 interaction in the $NN$ channel in panels (a) -- (c) and with the SkM$^*$ interaction in panels (d) -- (f). The $\Lambda N$ and $\Lambda\Lambda$ channels are described by the NSC97f+EmpC functional.}
\label{fig:dif_str}
\end{figure*}

The multipole strength distributions obtained with the SGII Skyrme interaction in the $NN$ channel, shown in Figs.~\ref{fig:monopole_full} -- \ref{fig:quadrupole_full}, exhibit a systematic upward energy shift with increasing $\Lambda$ hyperon number $-S$. To check that this behavior is not a specific consequence of the SGII parametrization, the calculations are repeated with other Skyrme-type interactions. In Fig.~\ref{fig:dif_str}(a) -- (c), the isoscalar monopole, isovector dipole, and isoscalar quadrupole strength distributions for $^{208-S}_{-S\Lambda}$Pb hypernuclei are shown for the SLy4 interaction, respectively. The corresponding results obtained with the SkM$^*$ interaction are presented in Fig.~\ref{fig:dif_str}(d) -- (f). It should be noted that, in Fig.~\ref{fig:dif_str}, the $\Lambda N$ and $\Lambda\Lambda$ interactions are described by the same NSC97f+EmpC functional as in Figs.~\ref{fig:monopole_full} -- \ref{fig:quadrupole_full}. As seen from Fig.~\ref{fig:dif_str}, for the $-S=0$ case, all calculated ISGMR, IVGDR, and ISGQR strength distributions of $^{208}$Pb lie in the corresponding experimentally established giant-resonance energy regions \cite{Tamii2011,RocaMaza2013,Patel2014}. For the finite-hyperon cases, Fig.~\ref{fig:dif_str} shows that the results obtained with both SLy4 and SkM$^*$ exhibit the same qualitative behavior as those calculated with the SGII interaction. In particular, for each multipole strength distribution, a systematic upward energy shift is observed with increasing strangeness number $-S$. This demonstrates that the hyperon-induced hardening of the collective modes is not a specific feature of the SGII parametrization, but a more general effect obtained also with other Skyrme interactions in the $NN$ channel.

\subsection{Giant Resonance Energies}

\begin{figure*}
\includegraphics[width=0.32\textwidth]{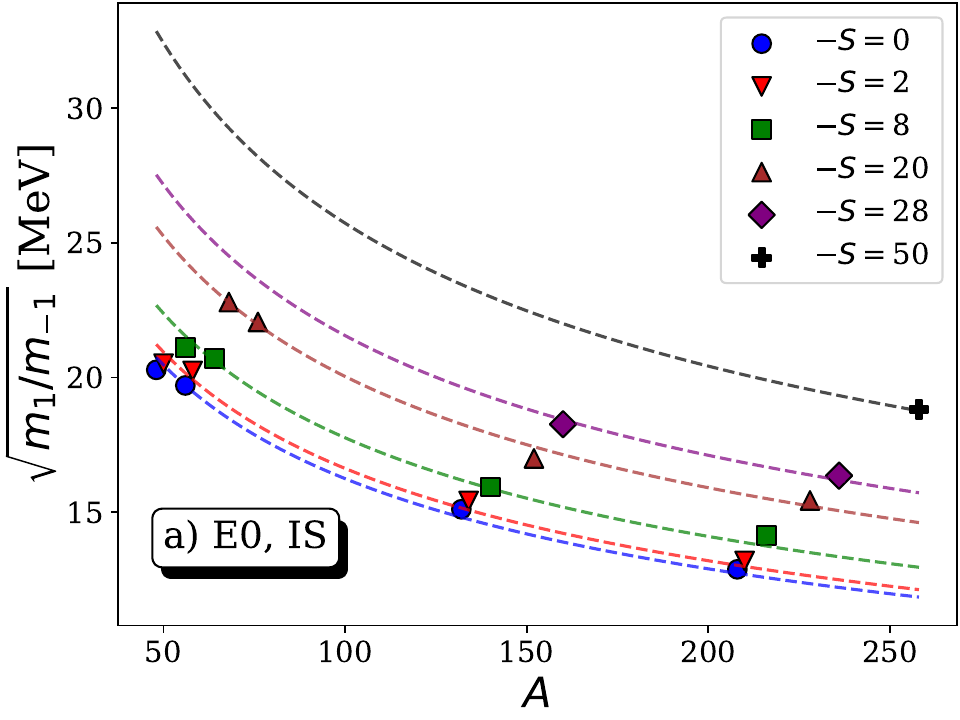}
\hfill
\includegraphics[width=0.32\textwidth]{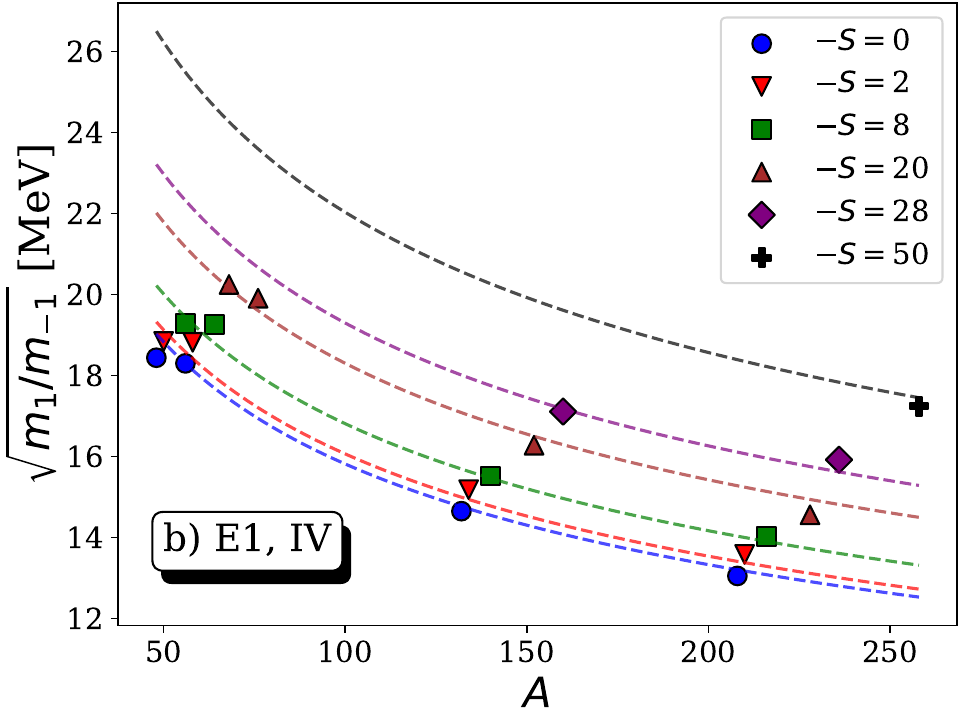}
\hfill
\includegraphics[width=0.32\textwidth]{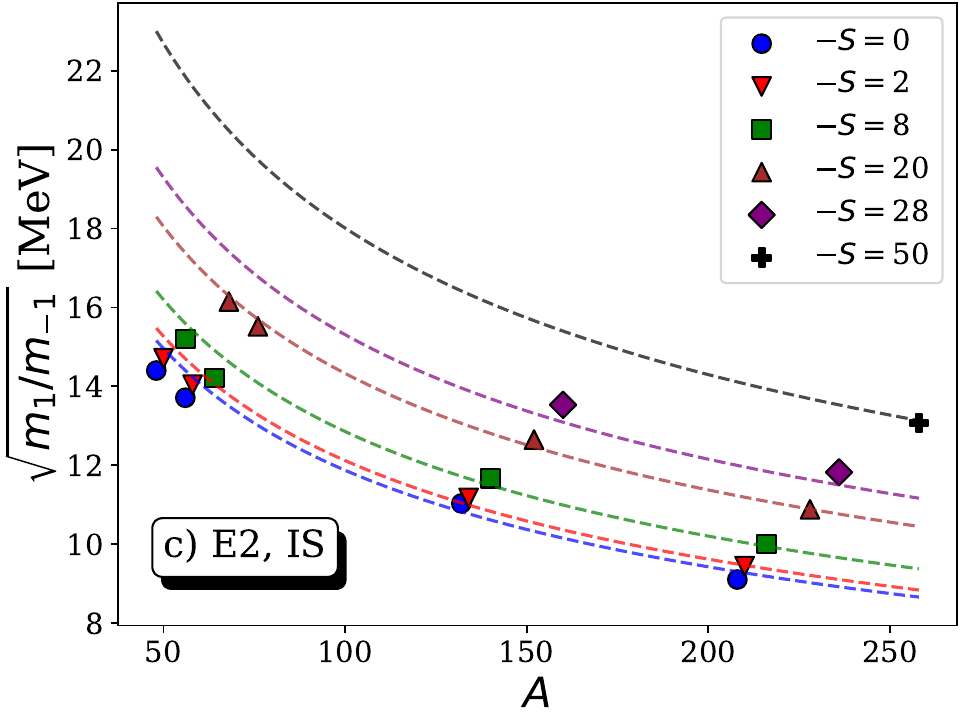}
\caption{Evolution of the centroid energies $\sqrt{m_1/m_{-1}}$ as a function of the mass number $A$ for the 
(a) isoscalar monopole (E0), (b) isovector dipole (E1), and (c) isoscalar quadrupole (E2) collective modes in 
$^{48-S}_{-S\Lambda}$Ca, $^{56-S}_{-S\Lambda}$Ni, $^{132-S}_{-S\Lambda}$Sn, and $^{208-S}_{-S\Lambda}$Pb hyperisotopes, calculated using the SGII-NSC97f+EmpC functional. 
Each value of strangeness $-S$ is depicted by a distinct color. The dashed lines represent the fitted scaling functions, 
given by (a) $\sqrt{m_1/m_{-1}} = 75.37 A^{-1/3} \left(1-0.01 S\right)$, (b) $\sqrt{m_1/m_{-1}} = \left(36.35 A^{-1/3} + 
17.19 A^{-1/6}\right) \left(1-0.008 S\right)$, and (c) $\sqrt{m_1/m_{-1}} = 55.12 A^{-1/3} \left(1-0.01 S\right)$.}
\label{fig:GR} 
\end{figure*}

The systematic upward shift of the strength distributions observed in Figs.~\ref{fig:monopole_full}, 
\ref{fig:dipole_full}, and \ref{fig:quadrupole_full} indicates a hardening of the nuclear equation of 
state. To quantify this effect, we calculated the centroid energies $\sqrt{m_1/m_{-1}}$ for the monopole, 
dipole, and quadrupole modes across the entire mass range from $^{48}_{0\Lambda}$Ca to $^{258}_{50\Lambda}$Pb. 
The calculated centroid energies are presented in Fig.~\ref{fig:GR}. The addition of 
$\Lambda$ hyperons to the system systematically shifts the centroid energies to higher values for all studied 
collective modes. By fitting these results, we find that the standard hydrodynamic scaling laws -- specifically 
the $A^{-1/3}$ dependence for the isoscalar monopole and quadrupole modes \cite{Blaizot1980, Speth1981}, and the 
mixed $A^{-1/3}$ and $A^{-1/6}$ dependence for the isovector dipole mode \cite{Berman1975} -- are preserved in 
multi-$\Lambda$ hypernuclei when modified by a linear stiffening factor $1-\gamma S$. Here, $\gamma$ represents 
the hyperon stiffening coefficient, a dimensionless parameter that quantifies the fractional increase in the 
resonance energy per unit of strangeness. 

The fitted functions are depicted by dashed lines in Fig.~\ref{fig:GR}. For the isoscalar monopole mode 
(Fig.~\ref{fig:GR}a), the scaling behavior is well described by $\sqrt{m_1/m_{-1}} = 75.37 A^{-1/3} \left(1-0.01 S\right)$, 
while for the quadrupole mode (Fig.~\ref{fig:GR}c), it follows $\sqrt{m_1/m_{-1}} = 55.12 A^{-1/3} \left(1-0.01 S\right)$. 
In the case of the isovector dipole mode (Fig.~\ref{fig:GR}b), the energy evolves as $\sqrt{m_1/m_{-1}} = 
\left(36.35 A^{-1/3} + 17.19 A^{-1/6}\right) \left(1-0.008 S\right)$. It is worth noting that for all considered collective 
modes, these three scaling functions reproduce the centroid energies with a relative error of less than 2\% across the mass 
table. These results demonstrate that the dynamical effect of strangeness is universal across the mass table: while the 
increasing mass $A$ tends to lower the collective frequency, the presence of strange baryons introduces a countervailing 
hardening effect that is linear with the number of $\Lambda$ hyperons. The typical stiffening factor $\gamma$ value, of 
about 1\%, shows that strangeness effect on the giant resonances is expected when the number of hyperon is typically larger that ten.

\subsection{Nuclear Incompressibility \label{sub:incom}}

\begin{figure}
\includegraphics[width=0.48\textwidth]{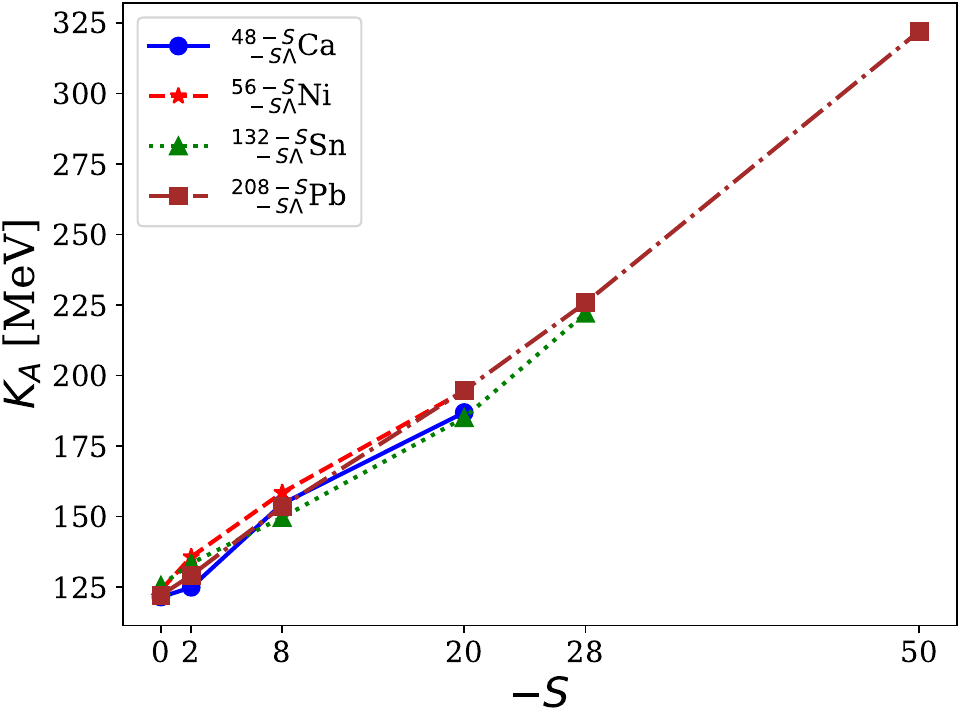}
\caption{Calculated nuclear incompressibility modulus $K_A$ as a function of the number of $\Lambda$ hyperons in 
$^{48-S}_{-S\Lambda}$Ca, $^{56-S}_{-S\Lambda}$Ni, $^{132-S}_{-S\Lambda}$Sn, and 
$^{208-S}_{-S\Lambda}$Pb hyperisotopes, using the SGII+NSC97f+EmpC functional.}
\label{fig:KA_E0}
\end{figure}

The calculation of the centroid energy enables us to determine the nuclear incompressibility of the studied hypernuclei. 
This can be performed using the moment-based relation given below \cite{Blaizot1980, Minato2012}
\begin{equation}\label{eq:KA}
K_A = \frac{m_N m_\Lambda}{\hbar^2} \frac{A \langle r^2 \rangle^2 \frac{m_1}{m_{-1}}}{m_\Lambda(N+Z)\langle r^2 
\rangle_{\mathrm{n}+\mathrm{p}}-m_N S \langle r^2 \rangle_{\Lambda}},
\end{equation}
where $m_N$ and $m_{\Lambda}$ are the nucleon and $\Lambda$ masses 
respectively, and $\langle r^2 \rangle$ denotes the mean square radii for the various particle distributions.
By substituting the centroid energy results for $\sqrt{m_1/m_{-1}}$ (shown in Fig.~\ref{fig:GR}a) into Eq.~(\ref{eq:KA}), 
we can determine the evolution of the nuclear incompressibility modulus, $K_A$, as a function of the number of $\Lambda$ 
hyperons for the $^{48-S}_{-S\Lambda}$Ca, $^{56-S}_{-S\Lambda}$Ni, $^{132-S}_{-S\Lambda}$Sn, and $^{208-S}_{-S\Lambda}$Pb 
isotopes. The resulting values are presented in Fig.~\ref{fig:KA_E0}. From this figure, we observe that in the limit of 
ordinary nuclei ($S=0$), the extracted incompressibility values are clustered within a narrow range of 121–126~MeV. 
Specifically, we find $K_A=121.49$~MeV for $^{48}$Ca, 124~MeV for $^{56}$Ni, 126~MeV for $^{132}$Sn, and 122~MeV for $^{208}$Pb. 
As the strangeness content increases, a systematic enhancement in $K_A$ is observed for all isotopes. For 
instance, at $-S=8$, the incompressibility rises to approximately 154–158~MeV for the lighter nuclei (Ca, Ni) and 
150–153~MeV for the heavier ones (Sn, Pb). By $-S=20$, the $K_A$ values approach the range of 185–195~MeV. Notably, 
at this stage, the values for $^{76}_{20\Lambda}$Ni and $^{228}_{20\Lambda}$Pb (195 MeV) are nearly identical. For the heavier systems, 
this linear increasing trend continues. In $^{160}_{28\Lambda}$Sn, $K_A$ reaches 222~MeV at $-S=28$. For the heaviest 
system, $^{258}_{50\Lambda}$Pb, the calculations extend up to $-S=50$, where the incompressibility reaches its maximum 
calculated value of 322~MeV.
These results collectively indicate that, in the case of finite hypernuclei, the HFRPA calculations yield a
monotonic increase in $K_A$ as a function of the strangeness number $-S$.
 
\subsection{Hypernuclear matter at nuclear saturation density}

To show that the results obtained in Sec.~\ref{sub:incom} are driven by bulk properties of the system, we now calculate the 
incompressibility of uniform hypernuclear matter. This approach allows us to isolate the bulk effect of the strong interaction from
finite-size effects, such as the nuclear surface and Coulomb repulsion. In this uniform limit, we define the curvature matrix as \cite{Chomaz2004}
\begin{equation} \label{eq:Kmatrix}
\mathcal{K} = 9 \left(\begin{array}{cc}
\rho_N \frac{\partial^2 \epsilon}{\partial \rho_N^2} & \sqrt{\rho_N \rho_\Lambda} \frac{\partial^2 \epsilon}{\partial \rho_N 
\partial \rho_{\Lambda}} \\
\sqrt{\rho_N \rho_\Lambda} \frac{\partial^2 \epsilon}{\partial \rho_N \partial \rho_{\Lambda}} 
& \rho_\Lambda \frac{\partial^2 \epsilon}{\partial \rho_{\Lambda}^2}
\end{array}\right),
\end{equation}
where $\epsilon$, see Eq.~\eqref{eq:total_energy}, denotes the energy density of the system. In the case of symmetric nucleonic matter, where
$\rho_n = \rho_p = \rho_N / 2$, $\epsilon$ can be expressed as follows
\begin{eqnarray}
\epsilon(\rho_N,\rho_\Lambda)
&=& \frac{3\hbar^2}{10 m_N} \left(\frac{3 \pi^2}{2}\right)^{2/3} \rho_N^{5/3} + \frac{3}{8} t_0 \rho^2_N 
+ \frac{1}{16} t_3 \rho_N^{\alpha+2}\nonumber\\
&& \hspace{-1.5cm}+ \frac{3}{80} \left(\frac{3 \pi^2}{2}\right)^{2/3} \left(3t_1 
+ t_2 \left(5+4x_2\right)\right)\rho_N^{8/3}\nonumber\\
&& \hspace{-1.5cm}+ \frac{3\hbar^2}{10 m_\Lambda} \left(3 \pi^2\right)^{2/3} \rho_\Lambda^{5/3} 
- \left(\alpha_1 - \alpha_2 \rho_N +\alpha_3 \rho_N^2\right) \rho_N \rho_\Lambda \nonumber\\
&& \hspace{-1.5cm}+ \left(\alpha_4 - \alpha_5 \rho_N +\alpha_6 \rho_N^2\right) \rho_N \rho_\Lambda^{5/3} 
- \left(\alpha_7 - \alpha_8 \rho_\Lambda +\alpha_9 \rho_\Lambda^2\right) \nonumber\\
&& \hspace{-1.5cm}\times \rho_\Lambda^2.
\end{eqnarray}
The eigenvalues and eigenvectors of this two-component system are denoted $K_\pm$ and $\psi_\pm$, respectively. They satisfy the 
standard eigenvalue equation $\mathcal{K} \psi_\pm = K_\pm \psi_\pm$. By diagonalizing $\mathcal{K},$ we can extract these 
eigenvalues and eigenvectors. From the characteristic equation $\det(\mathcal{K} - K_\pm \text{Id}) = 0$, we find that

\begin{figure}
\includegraphics[width=0.48\textwidth]{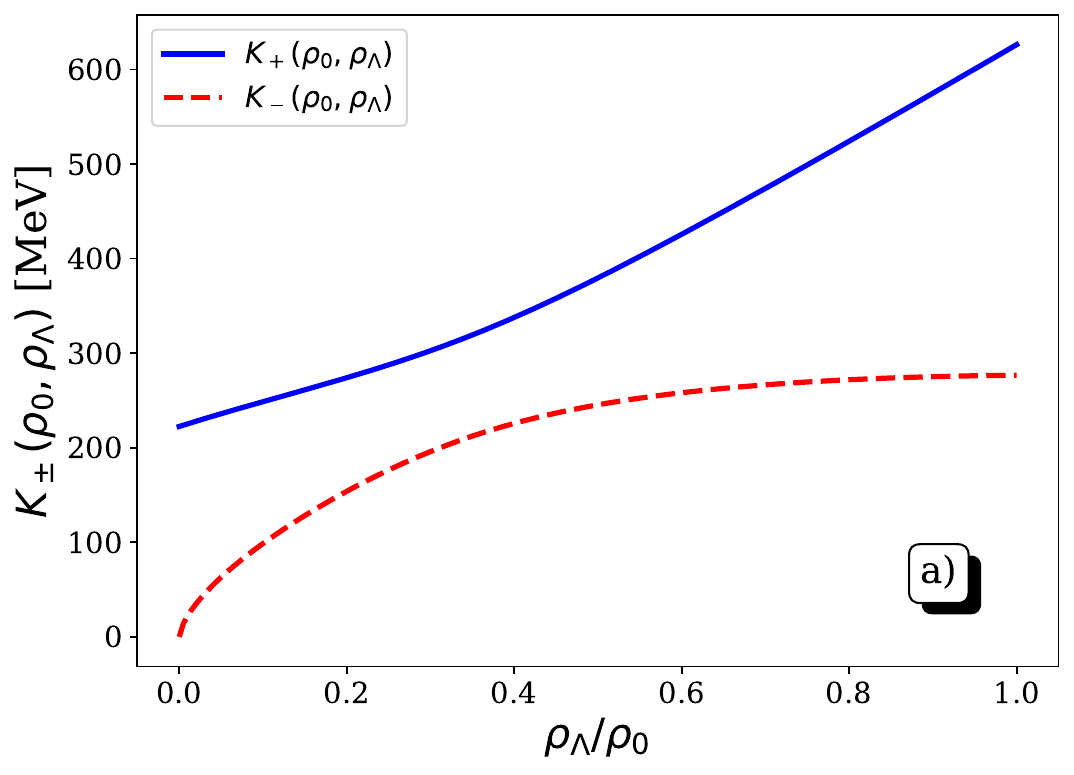}
\hfill
\includegraphics[width=0.48\textwidth]{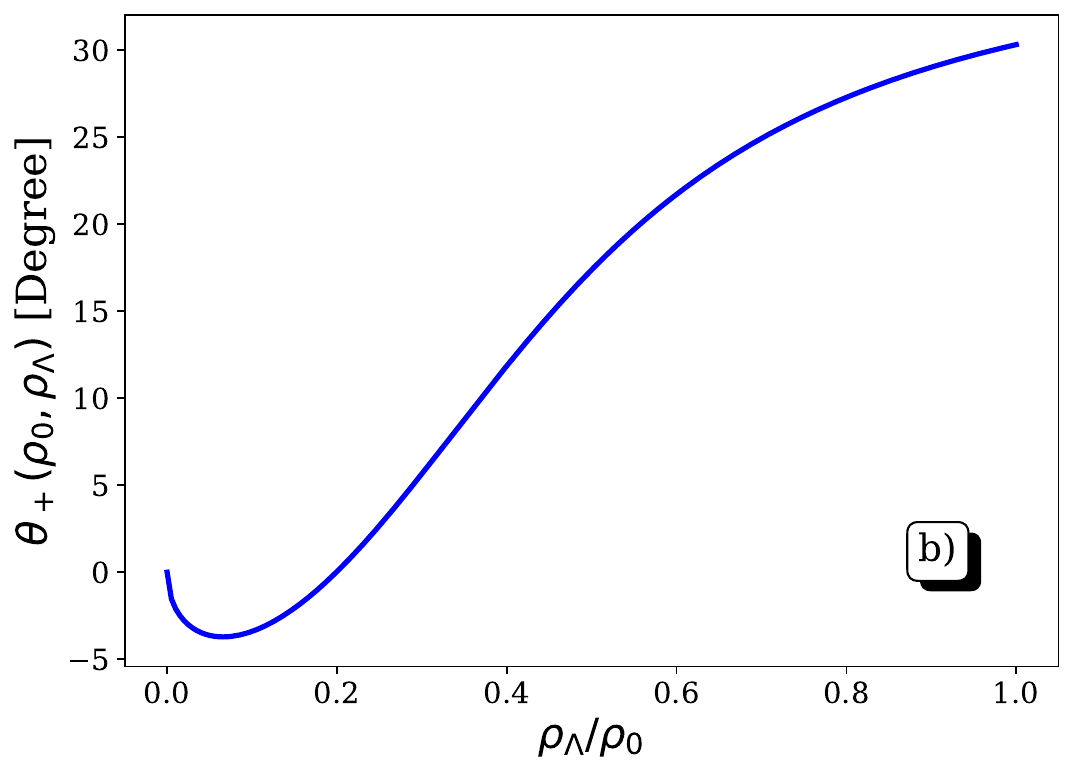}
\caption{Evolution of the curvature matrix eigenvalues ($K_\pm$) and the mixing angle ($\theta_+$) as a function of the scaled
$\Lambda$ density in uniform hypernuclear matter, evaluated at constant nucleon saturation density ($\rho_N = 
\rho_0 = 0.16$ fm$^{-3}$) using the SGII+NSC97f+EmpC functional.}
\label{fig:K_matter}
\end{figure}

\begin{widetext}
\begin{equation} \label{eq:eigen_incom_uni}
K_{\pm} (\rho_N, \rho_\Lambda) = \frac{9}{2} \left( 
\rho_N \frac{\partial^2 \epsilon}{\partial \rho_N^2} + \rho_\Lambda\frac{\partial^2 \epsilon}{\partial \rho_\Lambda^2} 
\pm \sqrt{\left(\rho_N \frac{\partial^2 \epsilon}{\partial \rho_N^2} + \rho_\Lambda 
\frac{\partial^2 \epsilon}{\partial \rho_\Lambda^2} \right)^2 - 4 \rho_N \rho_\Lambda\left(\frac{\partial^2 \epsilon}{\partial \rho_N^2} 
\frac{\partial^2 \epsilon}{\partial \rho_\Lambda^2} - \left(\frac{\partial^2 \epsilon}{\partial \rho_N \partial \rho_\Lambda}\right)^2\right)}
\right).
\end{equation}
\end{widetext}
Fig.~\ref{fig:K_matter}a shows the variation of the $K_{\pm}$ eigenvalues as a function of the 
scaled $\Lambda$ density in uniform hypernuclear matter, evaluated at constant nucleon saturation density ($\rho_N = 
\rho_0 = 0.16$ fm$^{-3}$). This figure shows that $K_{+} (\rho_0, \rho_\Lambda)$ is the dominant eigenvalue, and it reproduces the $\rho_\Lambda \rightarrow 0$ limit, where it is equal to the incompressibility of nucleonic matter. $K_{+} (\rho_0, \rho_\Lambda)$ represents the stiffest compressional mode of hypernuclear
matter. The monotonic increase of $K_{+} (\rho_0, \rho_\Lambda)$ confirms that hypernuclear matter becomes stiffer as a function of the increase of the strangness number $-S$, which is also observed in hypernuclei via HFRPA calculations, see Fig.~\ref{fig:KA_E0}. Therefore, the stiffening predicted in the finite system is a consequence of the bulk $N\Lambda$ and $\Lambda\Lambda$ interactions, and not merely an artifact of the finite volume in nuclei. It should be 
noted that Fig.~\ref{fig:K_matter}a also shows the variation of the eigenvalue $K_{-} (\rho_0, \rho_\Lambda)$ with respect to $\rho_\Lambda$. $K_{-} (\rho_0, \rho_\Lambda)$ represents the second eigenvalue, corresponding to the hyperon-dominated mode that emerges dynamically as $\Lambda$ particles are introduced into 
the system. This eigenvalue vanishes in the limit $\rho_\Lambda \rightarrow 0$ as expected.

\begin{figure}
\includegraphics[width=0.48\textwidth]{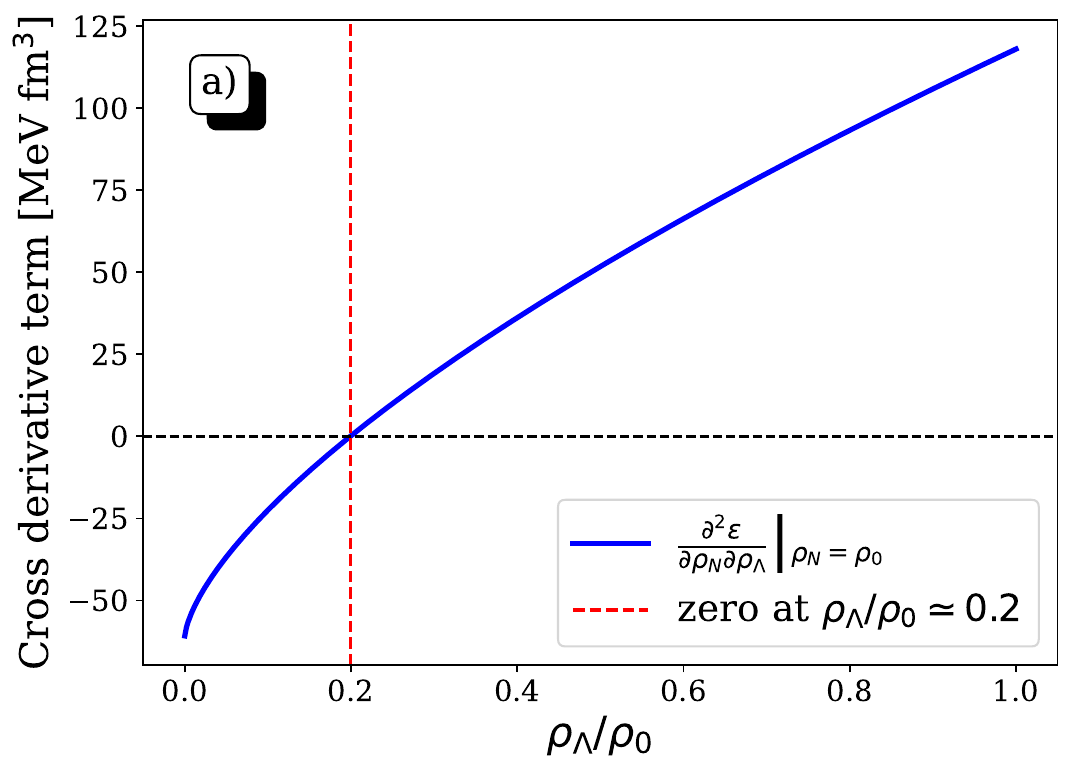}
\hfill
\includegraphics[width=0.48\textwidth]{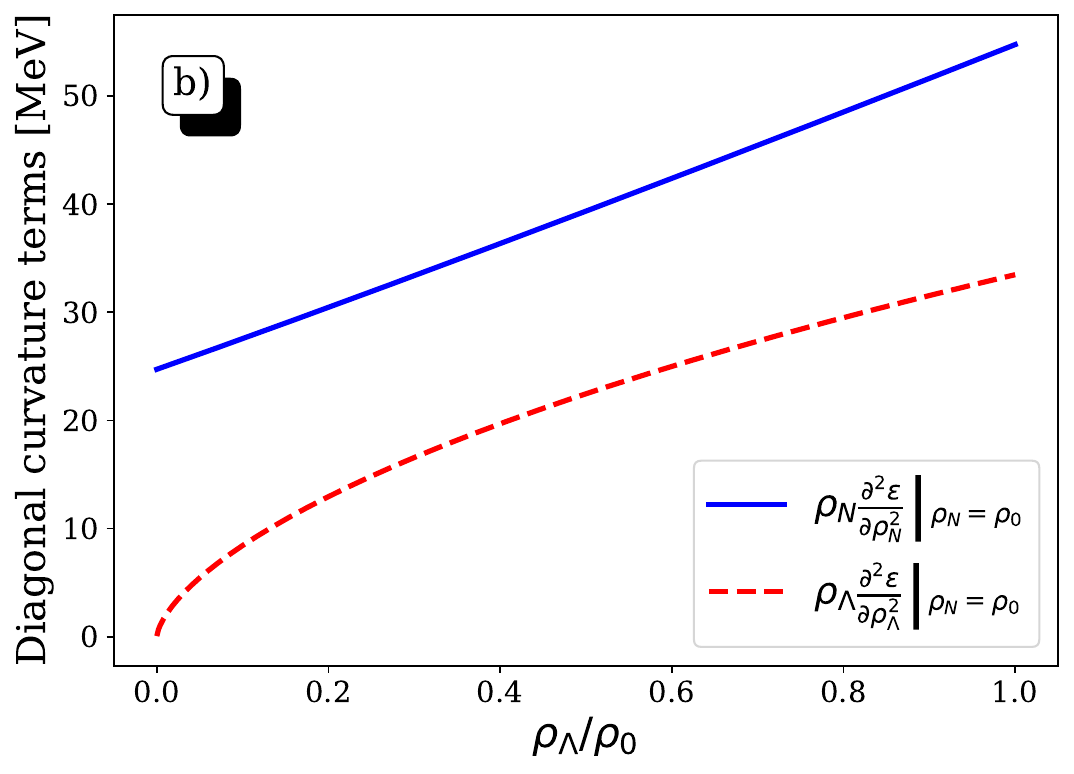}
\caption{Variation of the cross derivative 
$\left.\frac{\partial^2\epsilon}{\partial\rho_N\partial\rho_\Lambda}\right|_{\rho_N=\rho_0}$ (a), and the diagonal curvature terms 
$\left.\rho_N\frac{\partial^2\epsilon}{\partial\rho_N^2}\right|_{\rho_N=\rho_0}$ and 
$\left.\rho_\Lambda\frac{\partial^2\epsilon}{\partial\rho_\Lambda^2}\right|_{\rho_N=\rho_0}$ (b), as functions of $\rho_\Lambda/\rho_0$, calculated using the SGII+NSC97f+EmpC functional.}
\label{fig:matter_terms}
\end{figure}

To understand how the nucleonic and hyperonic density fluctuations influence each other, we also examine the eigenvectors
of the curvature matrix, given in Eq.~(\ref{eq:Kmatrix}). Because this matrix is symmetric, the eigenvectors can be parameterized as follows
\begin{equation}
\psi_\pm = \vert \psi_\pm \vert\binom{\cos \theta_\pm}{\sin \theta_\pm},
\end{equation}
where $\theta_\pm$ is an angle informing about the direction of the mode in the nucleonic and hyperonic space. This angle represents the degree of mixing between the purely nucleonic and purely hyperonic compressional modes.  For each of the eigenmodes, the angle $\theta_\pm$ determines the degree of
mixing between the nucleonic and $\Lambda$ density fluctuations. The in-phase
or out-of-phase character of the mode is then determined by the relative sign
of the nucleonic and hyperonic components of the corresponding eigenvector.
The density-dependent mixing angle $\theta_+ (\rho_N, \rho_\Lambda)$ is 
explicitly given by 
\begin{equation}\label{eq:angle}
\theta_+ (\rho_N, \rho_\Lambda) =\frac{1}{2} \arctan \left(\frac{2 \sqrt{\rho_N \rho_\Lambda} \frac{\partial^2 \epsilon}{\partial \rho_N 
\partial \rho_{\Lambda}}}{\rho_N \frac{\partial^2 \epsilon}{\partial \rho_N^2}-\rho_\Lambda \frac{\partial^2 \epsilon}{\partial \rho_\Lambda^2}}\right).
\end{equation}
Fig.~\ref{fig:K_matter}b illustrates the evolution of this mixing angle as a function of the $\Lambda$ density. As seen in this figure, $\theta_+$ is first negative and then positive. The origin of this change of sign lies in the cross derivative term in the numerator of the angle equation (\ref{eq:angle}).
Evaluating the cross derivative term at $\rho_N=\rho_0$, we obtain
\begin{equation}
\left.
\frac{\partial^2 \epsilon}
{\partial \rho_N \partial \rho_{\Lambda}}
\right|_{\rho_N=\rho_0}
=
-61.09 + 607.67\,\rho_\Lambda^{2/3},
\end{equation}
whose evolution as a function of $\rho_\Lambda/\rho_0$ is shown in
Fig.~\ref{fig:matter_terms}a. As seen from this figure, the cross derivative
is negative at very low hyperon densities
$\left(\rho_{\Lambda} / \rho_0<0.2\right)$. This behavior originates from the
dominance of the attractive constant term $-61.09$ over the repulsive
density-dependent term proportional to $\rho_\Lambda^{2/3}$. From
Fig.~\ref{fig:matter_terms}b, we furthermore observe that 
\begin{equation}
\left.
\rho_N\frac{\partial^2\epsilon}{\partial\rho_N^2}
\right|_{\rho_N=\rho_0}
>
\left.
\rho_\Lambda\frac{\partial^2\epsilon}{\partial\rho_\Lambda^2}
\right|_{\rho_N=\rho_0}
\end{equation}
throughout the entire interval $0 \le \rho_\Lambda/\rho_0 \le 1$.
As a consequence, the mixing angle $\theta_+$ becomes negative for
$\rho_\Lambda/\rho_0<0.2$, indicating that, within the present eigenvector
convention, the nucleonic and hyperonic components of the corresponding
eigenvector have opposite signs. However, as the hyperon density increases,
the positive density-dependent contribution proportional to
$\rho_\Lambda^{2/3}$ rapidly overtakes the attractive term. The cross
derivative then becomes positive, driving $\theta_+$ toward positive values.
In this high-density regime, the nucleonic and hyperonic components of the
corresponding eigenvector acquire the same sign and become strongly coupled,
which directly contributes to the stiffening observed in the eigenvalue
$K_{+}(\rho_0,\rho_\Lambda)$.

\subsection{Differences in transition densities induced by $\Lambda$ hyperons}

\begin{figure*}
\includegraphics[width=0.32\textwidth]{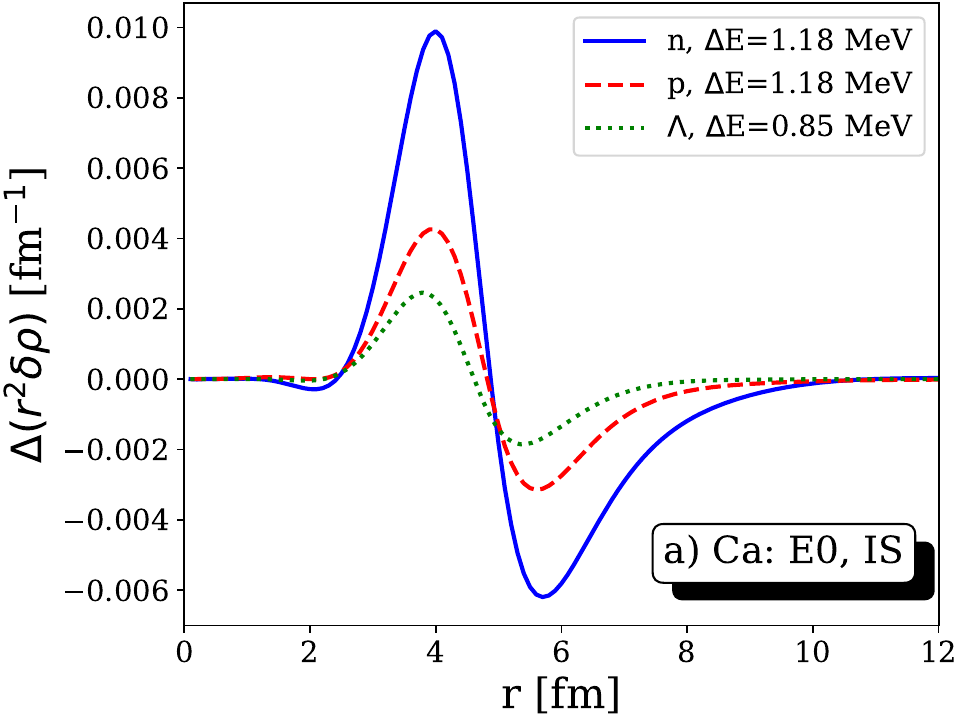}
\hfill
\includegraphics[width=0.32\textwidth]{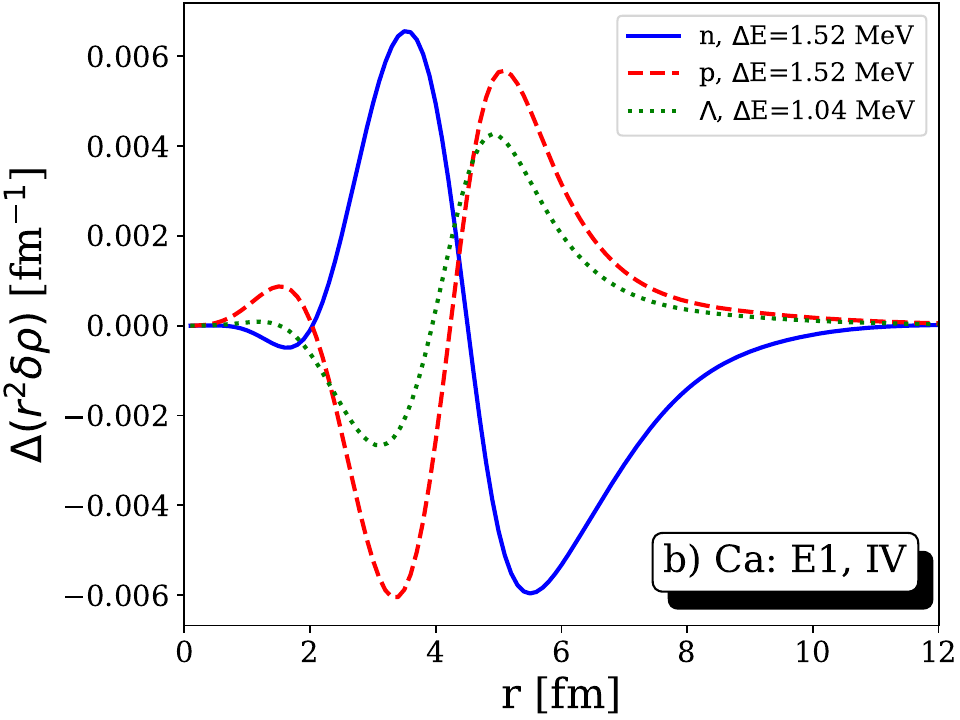}
\hfill
\includegraphics[width=0.32\textwidth]{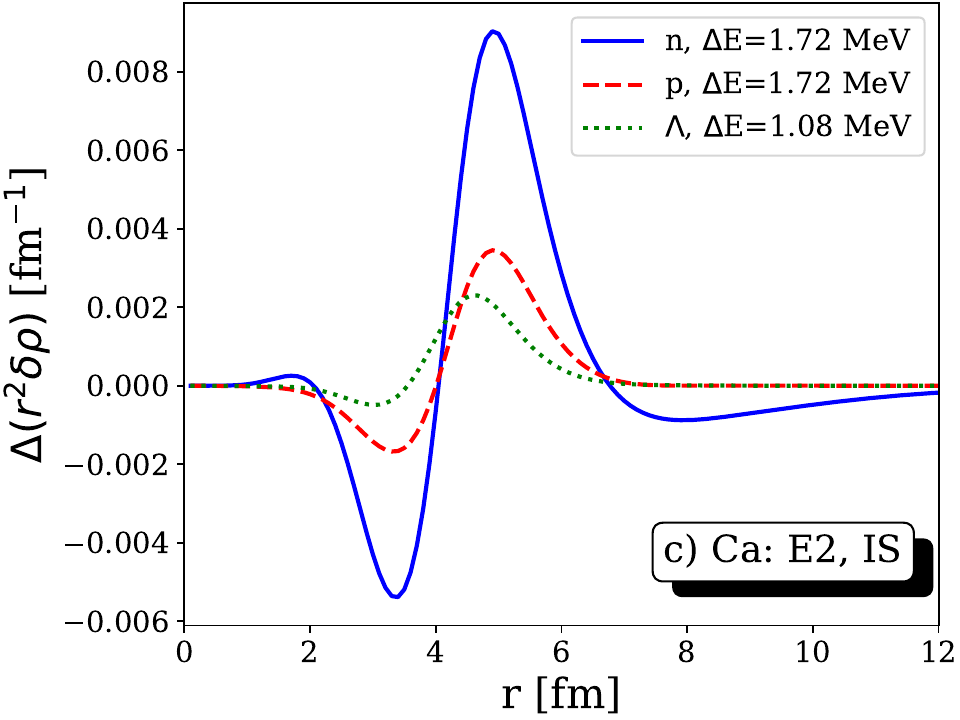}
\hfill
\includegraphics[width=0.32\textwidth]{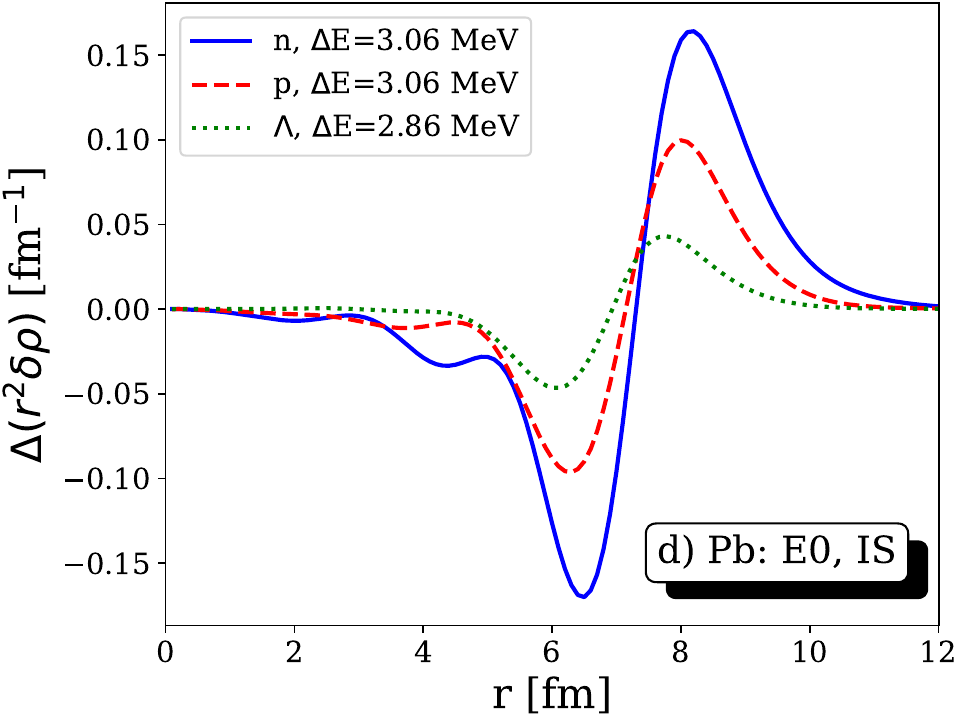}
\hfill
\includegraphics[width=0.32\textwidth]{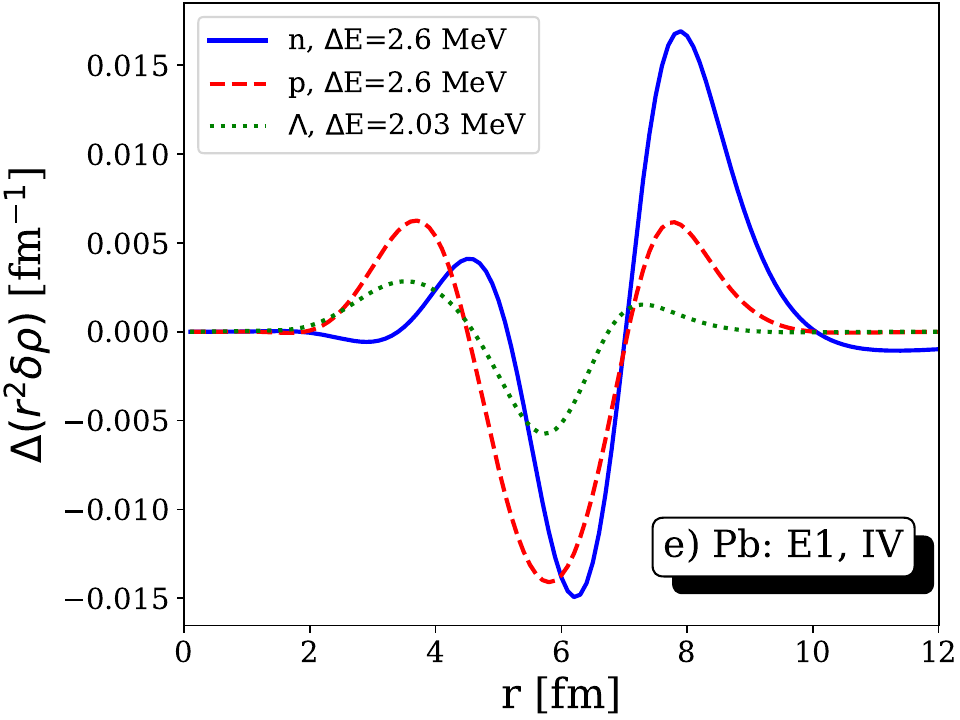} 
\hfill
\includegraphics[width=0.32\textwidth]{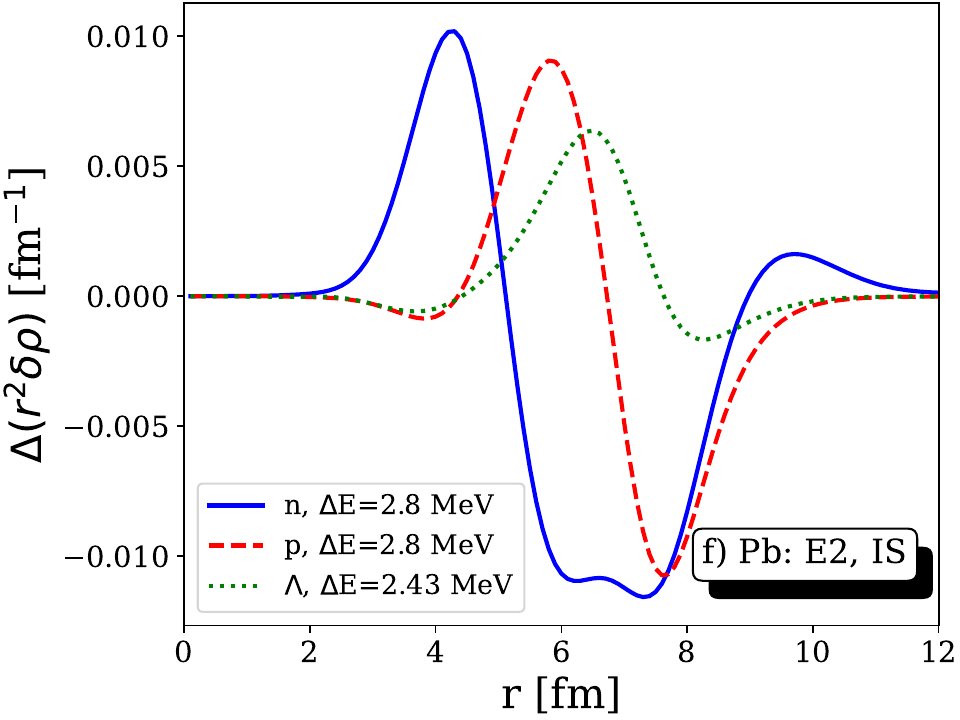} 
\caption{Differences in the transition densities weighted by the square of the radial coordinate, $\Delta(r^2\delta\rho)$, at the excitation energies corresponding to the giant resonance peaks in the $^{48-S}_{-S\Lambda}$Ca and $^{208-S}_{-S\Lambda}$Pb isotopic chains, calculated using the SGII+NSC97f+EmpC functional. Here, $\Delta E$ represents the difference between the corresponding excitation energies given in Table~\ref{tab:td_energies}. Scaling factors are $c=1$ for nucleons and $c=10$ for $\Lambda$ hyperons.}
\label{fig:dif_td_Ca_Pb}
\end{figure*}

To investigate how $\Lambda$ hyperons affect the spatial profile of the collective modes, we define the differences in the transition densities weighted by the square of the radial coordinate between the hypernuclei containing the largest and smallest numbers of $\Lambda$ hyperons in a given isotopic chain as
\begin{equation}
\Delta \left(r^2 \delta \rho\right)
=
c \left[
r^2 \delta \rho(^{N-S_{\max}}_{-S_{\max}}X)
-
r^2 \delta \rho(^{N-S_{\min}}_{-S_{\min}}X)
\right].
\end{equation}
The radial dependence of $\Delta \left(r^2 \delta \rho\right)$ at the excitation energies corresponding to the peaks of the giant resonances is shown in Fig.~\ref{fig:dif_td_Ca_Pb} for neutrons, protons, and $\Lambda$ hyperons in the $^{48-S}_{-S}$Ca and $^{208-S}_{-S}$Pb isotopic chains. Panels (a)--(c) and (d)--(f) correspond to the isoscalar monopole, isovector dipole, and isoscalar quadrupole states for the Ca and Pb isotopic chains, respectively. The scaling factor is taken as $c=1$ for neutrons and protons, while $c=10$ is used for $\Lambda$ hyperons in order to improve the visibility of the corresponding transition densities. For the neutron and proton transition densities, the differences $\Delta(r^2\delta\rho)$ are calculated between the hypernuclei with maximal and minimal strangeness, corresponding to $-S_{\min}=0$ and $-S_{\max}=20$ for the $^{48-S}_{-S}$Ca isotopic chain and $-S_{\max}=50$ for the $^{208-S}_{-S}$Pb isotopic chain. Since no $\Lambda$-hyperon transition density exists for $-S=0$, the $\Lambda$-hyperon differences are calculated using $-S_{\min}=2$. The excitation energies corresponding to the hypernuclei employed in the present calculations are listed in Table~\ref{tab:td_energies}.
\begin{table}[t]
\caption{Excitation energies (in MeV) corresponding to the giant resonance peaks used in the transition-density calculations for selected hypernuclei in the $^{48-S}_{-S}$Ca and $^{208-S}_{-S}$Pb isotopic chains, calculated using the SGII+NSC97f+EmpC functional.}
\begin{ruledtabular}
\begin{tabular}{cccc}
Hypernucleus & $E0$ (IS) & $E1$ (IV) & $E2$ (IS) \\
\hline
$^{48}\mathrm{Ca}$ & 19.35 & 18.76 & 17.53 \\
$^{50}_{2\Lambda}\mathrm{Ca}$ & 19.68 & 19.24 & 18.17 \\
$^{68}_{20\Lambda}\mathrm{Ca}$ & 20.53 & 20.28 & 19.25 \\
\hline
$^{208}\mathrm{Pb}$ & 13.55 & 13.41 & 11.17 \\
$^{210}_{2\Lambda}\mathrm{Pb}$ & 13.75 & 13.98 & 11.54 \\
$^{258}_{50\Lambda}\mathrm{Pb}$ & 16.61 & 16.01 & 13.97 \\
\end{tabular}
\end{ruledtabular}
\label{tab:td_energies}
\end{table}
Fig.~\ref{fig:dif_td_Ca_Pb} shows that, for the monopole, dipole, and quadrupole states, the influence of $\Lambda$ hyperons on the transition densities becomes most pronounced in the radial region $r>2$~fm. In general, for all collective modes, the neutron transition densities exhibit the strongest modification induced by the presence of $\Lambda$ hyperons, whereas the changes in the $\Lambda$-hyperon transition densities remain comparatively weak. 

As observed in panels (b) and (e) of Fig.~\ref{fig:dif_td_Ca_Pb}, 
which correspond to the isovector dipole state, the transition densities 
of the protons and the $\Lambda$ hyperons oscillate in phase. 
In the Ca case, this motion is predominantly out of phase with respect to 
the neutron transition density, while in Pb the neutron response shows a more 
complex radial structure. This behavior indicates that, although $\Lambda$ hyperons are electrically neutral and insensitive to the Coulomb interaction in a neutron-rich system, they remain dynamically coupled to the proton fluid through the nucleon-hyperon interaction during the collective oscillation. This may be qualitatively understood from the fact that the proton number $Z$ is closer to the hyperon number $-S$ than to the neutron number $N$. Therefore, 
the global single-particle state filling are more similar between protons and hyperons, than between neutrons and hyperons. 

\section{\label{sec:concs} Conclusions}

In this study, we presented comprehensive Hartree-Fock Random Phase Approximation calculations 
of multipole strength distributions for multi-$\Lambda$ hypernuclei, focusing specifically on the spherical
isotopes of Ca, Ni, Sn, and Pb. Employing the Skyrme-type SGII interaction in the nucleon-nucleon channel and the 
NSC97f interaction in the hyperon channel, we systematically examined how the inclusion of $\Lambda$ hyperons modifies the
excited states of the nuclear core.

Our analysis of the strength distributions reveals a universal feature where the ISGMR, IVGDR, and ISGQR
all exhibit a systematic upward energy shift with increasing strangeness number $-S$. While the addition of 
$\Lambda$ hyperons deepens the central potential and contracts the nuclear density, uniform matter 
calculations predict that this hardening is a bulk effect 
driven by the strong $N\Lambda$ and $\Lambda\Lambda$ interactions.

While the upward energy shift is universal, the modification of the resonance shape differs among the modes. The
monopole response exhibits significant fragmentation and redistribution of spectral weight. In contrast, the 
quadrupole and dipole resonances largely
retain their collective peak structures and shapes, despite the substantial energy shifts induced by the large content of
strangeness.

We quantified the scaling behavior of the centroid energies $\sqrt{m_1/m_{-1}}$ as a function of mass number $A$ and 
strangeness $-S$. The scaling behavior of the computed centroid energies is well reproduced by the relations 
$\sqrt{m_1/m_{-1}} = 75.37 A^{-1/3} \left(1-0.01 S\right)$ for the isoscalar 
monopole mode, $\sqrt{m_1/m_{-1}} = \left(36.35 A^{-1/3} + 17.19 A^{-1/6}\right) \left(1-0.008 S\right)$ 
for the isovector dipole mode, and $\sqrt{m_1/m_{-1}} = 55.12 A^{-1/3} \left(1-0.01 S\right)$ for the 
isoscalar quadrupole mode. 

From the monopole energies, we extracted the nuclear incompressibility modulus $K_A$. We found a monotonic increase of 
$K_A$ with increasing $-S$ across all isotopic chains. The maximum incompressibility was identified for the $^{258}_{50\Lambda}$Pb
hypernucleus, reaching a value of $K_A=322$~MeV, 
in agreement with the behavior observed in hypernuclear matter. This shows that the increase of the incompressibility with the strangeness number $-S$ is a bulk effect originating in the behavior of uniform matter.

Finally, the analysis of the difference transition densities for the most collective states indicates that the dynamical influence of $\Lambda$ hyperons is predominantly localized in the radial region $r>2$~fm. Moreover, for the isovector dipole excitations, the proton and 
$\Lambda$-hyperon transition densities are found to oscillate 
in phase.


\begin{acknowledgments}
This work is supported by the Scientific and Technological Research Council of Turkey (T\"{U}B\.{I}TAK) under project number MFAG-125F501, MFAG-118F098, and the Yildiz Technical University under project number  FBA-2018-3325. J.M. and E.K. are supported by the CNRS-IN2P3 MAC2 masterproject and this work benefited from the support of the project RELANSE ANR-23-CE31-0027-01 of the French National Research Agency (ANR). 

\end{acknowledgments}


\newpage
\bibliographystyle{apsrev4-2}
\bibliography{z_bib}

\end{document}